\pdfoutput=1
\documentclass[aps,arxiv,twocolumn,superscriptaddress]{revtex4-1}
\usepackage{graphicx}
\usepackage{float}
\usepackage{amsmath}
\usepackage{amssymb}
\usepackage{mathtools}
\usepackage[utf8]{inputenc}
\usepackage[T1]{fontenc}
\usepackage[version=4]{mhchem}
\usepackage{booktabs}

\usepackage{microtype}
\usepackage[colorlinks, allcolors=blue]{hyperref}
\hypersetup{pdftitle={surface phase diagram of LiNiO2}, pdfauthor={X.~Li, A.~Urban}}

\newlength{\colwidth}
\begin{document}

\title{Understanding the onset of surface degradation in \ce{LiNiO2} cathodes}

\author{Xinhao Li}
\affiliation{Department of Chemical Engineering, Columbia University, New York, NY 10027, USA.}
\affiliation{Columbia Electrochemical Energy Center, Columbia University, New York, NY 10027, USA.}
\author{Qian Wang}
\affiliation{Department of Chemical Engineering, Columbia University, New York, NY 10027, USA.}
\author{Haoyue Guo}
\affiliation{Department of Chemical Engineering, Columbia University, New York, NY 10027, USA.}
\author{Nongnuch Artrith}
\affiliation{Department of Chemical Engineering, Columbia University, New York, NY 10027, USA.}
\affiliation{Columbia Center for Computational Electrochemistry, Columbia University, New York, NY 10027, USA.}
\affiliation{Materials Chemistry and Catalysis, Debye Institute for Nanomaterials Science, Utrecht University, 3584 CG Utrecht, The Netherlands}
\author{Alexander Urban}
\email{a.urban@columbia.edu}
\affiliation{Department of Chemical Engineering, Columbia University, New York, NY 10027, USA.}
\affiliation{Columbia Electrochemical Energy Center, Columbia University, New York, NY 10027, USA.}
\affiliation{Columbia Center for Computational Electrochemistry, Columbia University, New York, NY 10027, USA.}
\date{\today}

\begin{abstract}
   Nickel-based layered oxides offer an attractive platform for the
  development of energy-dense cobalt-free cathodes for lithium-ion
  batteries but suffer from degradation via oxygen gas release during
  electrochemical cycling.
  While such degradation has previously been characterized
  phenomenologically with experiments, an atomic-scale understanding of
  the reactions that take place at the cathode surface has been lacking.
  Here, we develop a first-principles methodology for the prediction of
  the surface reconstructions of intercalation electrode particles as a
  function of the temperature and state of charge.
  We report the surface phase diagrams of the \ce{LiNiO2(001)} and
  \ce{(104)} surfaces and identify surface structures that are
  likely visited during the first charge and discharge.
  Our calculations indicate that both surfaces experience oxygen loss
  during the first charge, resulting in irreversible changes to the
  surface structures.
  At the end of charge, the surface Ni atoms migrate into tetrahedral
  sites, from which they further migrate into Li vacancies during
  discharge, leading to Li/Ni mixed discharged surface phases.
  Further, the impact of the temperature and voltage range during
  cycling on the charge/discharge mechanism is discussed.
  The present study thus provides insight into the initial stages of
  cathode surface degradation and lies the foundation for the
  computational design of cathode materials that are stable against
  oxygen release.
\end{abstract}

\maketitle

\section{Introduction}
\label{sec:introduction}

The demand for lithium-ion batteries (LIBs) with greater capacity and energy density is rising, especially owing to the increasing adoption of electric vehicles, advancing penetration of renewable energy sources, and an increasing power consumption of portable electronics~\cite{s334-2011-928, ees5-2012-7854}.
The most energy-dense commercial LIBs employ cobalt-containing cathode materials, such as \ce{LiCoO2} (LCO) \cite{mrb15-1980-783, cm22-2010-587}, \ce{LiNi_xMn_{1-x/2}Co_{1-x/2}O2} with $x<1$ (NMC) \cite{cl30-2001-642, jps81-82-1999-416, aami4-2012-2329}, and \ce{LiNi_xCo_yAl_zO2} with $x>y\gg{}z$ (NCA) \cite{jops146-2005-594, joaac376-2004-282, aem8-2018-1703612}.
However, cobalt minerals are scarce, and most cobalt in global production is mined from the Democratic Republic of Congo (DRC)~\cite{m69-2017-37, ns1-2018-495, est54-2020-2985}.
The sustainable growth of LIB energy storage requires LIB chemistries that do not rely on scarce and expensive chemical elements~\cite{cr104-2004-4271, acie54-2015-4440} but are nevertheless energy-dense and mechanically, electrochemically, and thermally stable.

\ce{LiNiO2} (LNO) would, in principle, be a highly attractive alternative to \ce{LiCoO2} because the voltage of the \ce{Ni^{3+}}/\ce{Ni^{4+}} redox couple is slightly above \ce{Co^{3+}}/\ce{Co^{4+}}, and Ni is more abundant than Co, enabling potentially greater energy densities at lower cost~\cite{jes140-1993-1862, acie58-2019-10434, nc11-2020-1550, mt-2021-S1369702120304776}.
However, \ce{LiNiO2} (without other admixed metal species) exhibits properties that have so far prevented its commercialization as a LIB cathode material:
The synthesis of stoichiometric layered LNO is challenging, since \ce{Ni^{3+}} tends to disproportionate into \ce{Ni^{2+}} and \ce{Ni^{4+}} and the similar size of \ce{Ni^{2+}} and \ce{Li^+} makes cation mixing hard to avoid~\cite{mrb20-1985-1137, jes138-1991-2207, jes143-1996-1168, jopacos57-1996-1057}.
Additionally, even though the stability of lithiated LNO is comparable to LCO~\cite{ssi69-1994-6}, partially delithiated \ce{Li_{1-x}NiO2} is unstable and tends to decompose when heated even in air or inert gas~\cite{jota38-1992-295}.
This thermal instability of LNO is also a significant safety hazard, as it may result in the release of \ce{O2} gas and reactive oxygen species into the flammable electrolyte, potentially causing ignition~\cite{ssi109-1998-295, jes150-2003-A1450}.

LNO can be stabilized by admixing other metal species.
Co doping alone~\cite{ssi53-56-1992-681} does not prevent surface degradation~\cite{jes148-2001-A463}, but additionally introducing Al~\cite{jops128-2004-278} or Mn~\cite{jps81-82-1999-416, jps90-2000-176} significantly improves the stability of the material, and makes it suitable for LIB cathodes~\cite{jops119-121-2003-178, ael2-2017-196, jops233-2013-121, jes164-2017-A1361, jmca5-2017-874}.
However, even when substituted with Co and Mn or Al, Ni-rich cathode materials undergo thermal degradation following a similar mechanism as unmodified LNO that involves a surface phase transition from the layered structure (space group $R\overline{3}m$) to a disordered spinel structure (space group $Fd\overline{3}m$) and finally to a disordered rock-salt structure (space group $Fm\overline{3}m$)~\cite{cm23-2011-3953, cm26-2014-1084}.
Such surface degradation is further aggravated by Li excess, which has hampered the commercialization of Li- and Ni-rich cathodes such as Li-rich NMC~\cite{ees4-2011-2223, nl13-2013-3857}.

While the surface degradation of LNO and related cathode compositions have been characterized experimentally on a phenomenological level, a clear understanding of the reactions that take place at the cathode surface and the surface reconstructions that form on the atomic scale is still lacking.
Further insight into the atomic-scale processes that trigger surface degradation could potentially enable the design of modified cathode compositions that are more degradation resistant or protective coatings that can prevent surface reconstructions to occur.

First-principles calculations can offer insights into atomic-scale processes that are challenging to probe experimentally, and atomic-scale modeling has previously been used to investigate the degradation of LNO.
Das et al.\ computationally determined the bulk phase diagram of \ce{Li_{1-x}NiO_{2-y}} and the thermodynamics of phase transitions in the bulk of LNO upon Li extraction and oxygen gas release, identifying several phases that could be mistaken for spinel structures in diffraction experiments~\cite{cm29-2017-7840}.
Xiao et al.\ investigated the kinetics of the formation of densified phases near the surface of LNO~\cite{ael4-2019-811}.
Kong et al.\ applied a combination of first-principles calculations and experiments to the delithiation of LNO, finding evidence for oxygen redox participation that could result in oxygen release from the cathode surface~\cite{aem9-2019-1802586}.
The cathode surface was explicitly modeled in only a few studies.
Cho~et~al.\ investigated the surface stability and morphology of discharged \ce{LNO} with first-principles calculations, revealing that exposed surface oxygen will destabilize the surface and facilitate the oxidative decomposition of the electrolyte on the surface~\cite{aami9-2017-33257}.
Cheng~et~al.\ evaluated computationally the potential of different dopant species for enhancing the \ce{LNO} surface oxygen retention and validated with experiments that Sb~doping can improve the electrochemical performance~\cite{jmca-2020-10.1039.D0TA07706B}.
To our knowledge, no systematic theoretical study of oxygen release from a LIB cathode surface during cycling has been reported.

Here, we develop a general first-principles approach for the prediction of intercalation electrode surface phase diagrams as a function of the state of charge and the temperature.
The methodology is then applied to model oxygen release from the surface of LNO as a prototypical example to obtain a comprehensive picture of the surface reconstructions that may form during the initial stages of degradation.
The methodology is introduced in the following methods section~\ref{sec:methods}, predicted phase diagrams are reported in section~\ref{sec:results}, and the computational results are critically discussed in section~\ref{sec:discussion}.


\section{Methods}
\label{sec:methods}

The desired half reaction at a \ce{LiNiO2} positive electrode during charging, i.e., delithiation, is
\begin{align}
	\ce{Li_xNiO2 -> Li_{x-y}NiO2 + y Li^+ + y e^-}
	\label{eq:LNO-redox}
	\quad,
\end{align}
where $y$ moles of Li are extracted from the material, a corresponding amount of \ce{Ni^{3+}} oxidizes to \ce{Ni^{4+}}, and $x$ is close to 1 in the fully discharged state.
The partially delithiated \ce{LiNiO2} can undergo an undesired self-discharge reaction in which \ce{Ni^{4+}} is reduced to \ce{Ni^{3+}} and oxygen is oxidized, resulting in oxygen gas release:
\begin{align}
	\ce{Li_{x-y}NiO2 -> Li_{x-y}NiO_{2-z} + \frac{z}{2} O2 (^)}
	\label{eq:self-discharge}
	\quad.
\end{align}
This side reaction leads to the loss of lattice O, Li sites, and \ce{Ni^{3+/4+}} redox capacity, so that the discharge capacity is reduced compared to the charge capacity.
In response to the oxygen release, the electrode surface reconstructs, as the remaining atoms need to rearrange~\cite{nc5-2014-3529}.

To better understand the oxygen release mechanism and the nature of the surface reconstructions in \ce{LiNiO2}, we modeled different surface compositions and structures with systematically enumerated Li and O vacancies.
In a purely thermodynamic picture, the surface will reconstruct to form the structure with the lowest free energy at a given state of charge and temperature, providing an atomic-scale view of the degradation mechanism of \ce{LiNiO2}.

In the following, we describe the surface structure models, the defect enumeration approach, and the approximation of the surface free energy that underlie our methodology.

\begin{figure*}[t]
  \centering
  \includegraphics[width=\textwidth]{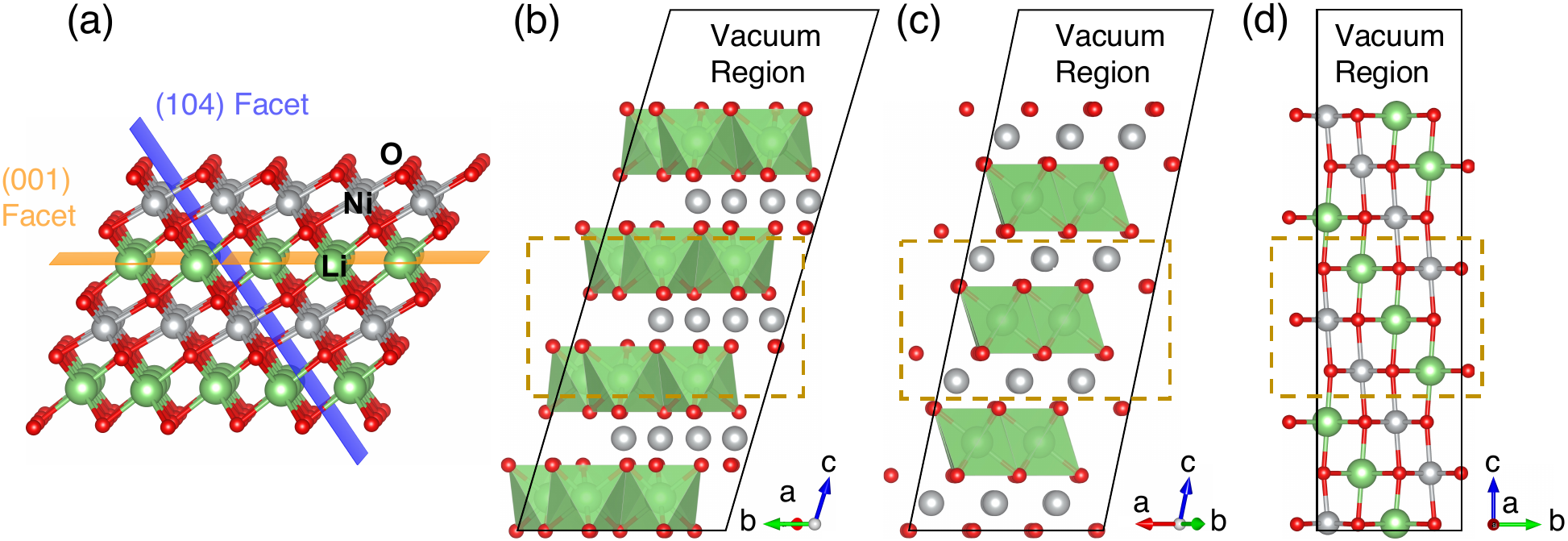}
  \caption{Visualizations of the surface slab models used as reference for the surface phase diagram calculations.  \textbf{(a)}~Section of the layered \ce{LiNiO2} structure highlighting the two investigated surface facets, \ce{(001) and \ce{(104)}}.  Periodic surface slab models of the \textbf{(b)}~Li-terminated and \textbf{(c)}~Ni-terminated \ce{LiNiO2(001)} facet as well as the \textbf{(d)}~non-polar $(104)$ facet.  The vacuum region of 15~\AA{} is truncated in the figures. Atomic planes that were held fixed at the ideal bulk sites are indicated with orange dashed rectangles, the remaining atoms were allowed to relax.  Li, Ni, and O atoms are shown in green, gray, and red, respectively.}
  \label{fig:slab-models}
\end{figure*}

\subsection{Atomic structure models of the \texorpdfstring{\ce{LiNiO2}}{LiNiO2} surfaces}
\label{sec:p1-structure-model}

The crystal structure of \ce{LiNiO2} consists of a cubic closed-packed (CCP) oxygen sublattice and a CCP cation sublattice in which the layers perpendicular to the $[001]$ direction (corresponding to the $[111]$ direction of the CCP sublattices) are occupied by alternating Li and Ni ions~\cite{cr104-2004-4271}.
In the pristine layered structure, Ni and Li ions are in octahedral sites (six-fold coordinated by oxygen).
At room temperature, the Ni sites in \ce{LiNiO2} exhibit a dynamic Jahn-Teller distortion~~\cite{cm32-2020-10096, prm4-2020-043601, cm30-2018-607}, and thus a fully optimized bulk structure model with distortion was used as the bulk reference in the present study.

Based on previous experimental and computational studies, the polar $(001)$ and the non-polar $(104)$ surfaces are the most stable and account for most of the exposed surface of \ce{LiNiO2}, \ce{LiCoO2}, and NMC cathode particles~\cite{cm21-2009-3799, aami9-2017-33257, jmca7-2019-5463, jmc22-2012-12874}.
Here, we therefore limit our analysis to these two \ce{LiNiO2} surface facets (\textbf{Figure~\ref{fig:slab-models}a}).
Note that the (104) facet intersects with the Li planes in the structure and thus permits Li extraction and intercalation.
The (001) facet is parallel to the cation layers, so that it is not directly relevant for intercalation but may still contribute to degradation modes.

\enlargethispage{\baselineskip}
Our approach makes use of periodic density-functional theory (DFT) calculations (described below) and therefore requires structure models with periodic symmetry also in the direction perpendicular to the modeled surfaces.
Truncating the \ce{LiNiO2} structure in the $[001]$ direction creates a surface structure with net dipole moment that cannot be stable in isolation~\cite{jpcssp12-1979-4977}.
However, in an electrochemical cell, the interaction with the environment, such as electrolyte molecules, may compensate surface dipoles, explaining why the unreconstructed $(001)$ facet is seen in experiments.
Our calculations generally employed surface slab models with inversion symmetry to remove any non-zero dipole moments from polar $(001)$ surface slab models.
All surface models were based on symmetric \ce{LiNiO2} slabs with 7~cation layers and a $2\times{}2$ surface unit cell.
The width of the vacuum region in the slab models was set to around 15~\AA{} following previous work~\cite{aami9-2017-33257}.
Li and Ni layers alternate in the $[001]$ direction, so that \ce{LiNiO2(001)} surface slab models can be either \ce{Li-O} or \ce{Ni-O} terminated.
We refer to these as \emph{Li-terminated} and \emph{Ni-terminated} slab models in the following.
Schematics of the base $(001)$ surface slab models are shown in \textbf{Figure~\ref{fig:slab-models}b--c}.

\begin{figure*}
	\centering
	\includegraphics[width=0.9\textwidth]{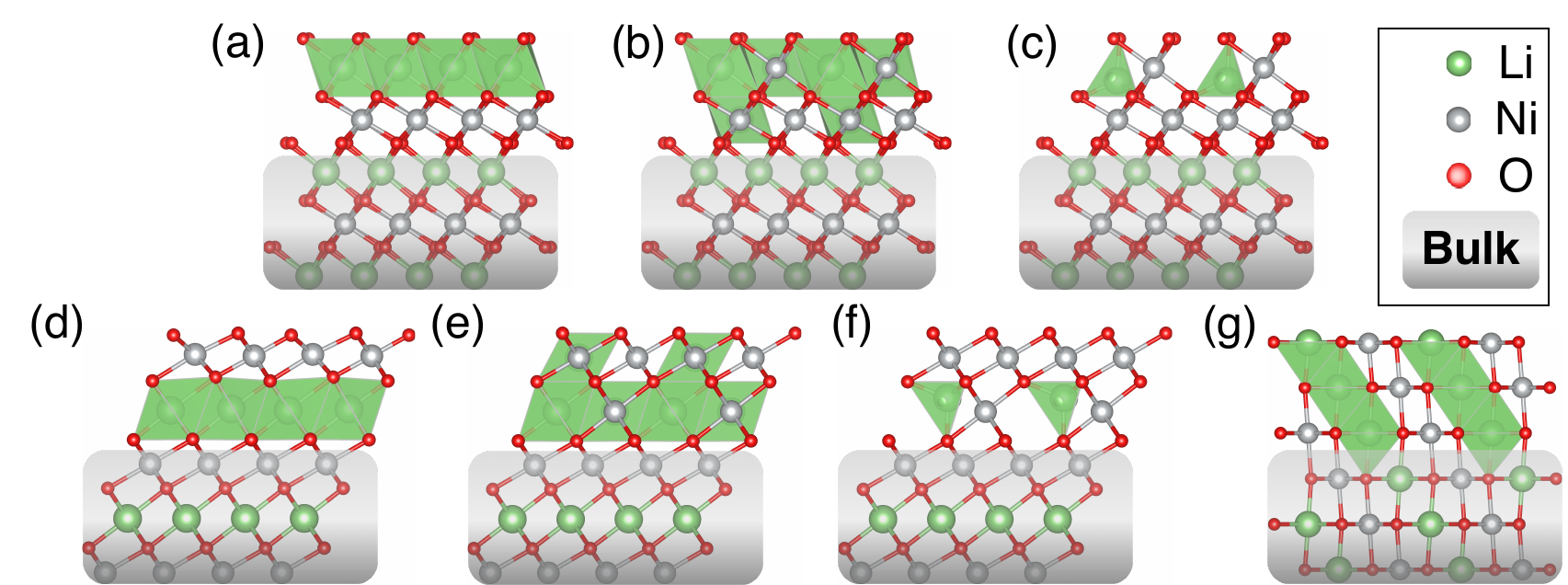}
	\caption{Active regions of the \textbf{(a--c)}~Li-terminated structure models based on the (a)~layered, (b)~spinel-like, and (c)~spinel structures.  \textbf{(d--f)}~Ni-terminated structure models based on the (d)~layered, (e)~spinel-like, and (f)~spinel structures.  \textbf{(g)}~Active region of the stoichiometric \ce{LiNiO2(104)} surface structure model.  The figures only show one side of the slab models, since the other sides are identical with inversion symmetry.}
	\label{fig:active-regions}
	\vspace{-\baselineskip}
\end{figure*}

The non-polar $(104)$ surface slab model can be treated as Tasker type~I surface~\cite{jpcssp12-1979-4977} because each layer contains the three species Li, Ni, and O in stoichiometric ratio and the surface does not exhibit a dipole moment.
Only one termination can be obtained for the non-polar $(104)$ surface, since all layers are equivalent (\textbf{Figure~\ref{fig:slab-models}d}).

In all surface slab calculations, the three central atomic layers were kept fixed at their ideal bulk coordinates to model the surfaces of extended materials.
The positions of the remaining atoms at the top and bottom of the slabs were fully optimized, and we refer to these as the \textit{active regions} of the slab models (see also \textbf{Figure~\ref{fig:slab-models}}).

In addition to surface models derived from the layered crystal structure, surface slabs with Li and Ni atoms rearranged as in the spinel crystal structure (space group $Fd\overline{3}m$) were also considered, since the spinel structure is known to be stable in the bulk when the material is half delithiated (\ce{LiNi_2O_4})~\cite{cm29-2017-7840}.
Two types of \emph{spinel-like} slab models of the $(001)$ surface were considered:
(i)~a slab model based on the over-lithiated \ce{Li_2Ni_2O_4} structure where Li and Ni ions are mixed but both reside in octahedral sites~(\textbf{Figure~\ref{fig:active-regions}b,e}), and (ii)~a slab model with Li ions in tetrahedral sites based on the true \ce{LiNi_2O4} spinel crystal structure~(\textbf{Figure~\ref{fig:active-regions}c,f}).
As for the layered $(001)$ surface slab models, the spinel-like $(001)$ slab models also have two different terminations, a mostly Li- and a mostly Ni-terminated surface.
Therefore, in total, 6~polar slab models and one non-polar slab model were considered.
The \emph{active regions} of the 7 slab models are shown in \textbf{Figure~\ref{fig:active-regions}}.

\subsection{Computational delithiation and enumeration of oxygen vacancies}
\label{sec:reconstruction-models}

The active regions of the surface structure models shown in \textbf{Figure~\ref{fig:active-regions}} were computationally delithiated, and for each lithium content oxygen vacancies were systematically enumerated.
The slab models with $2\times2$ surface unit cells contain four Li atoms in each lithium layer and four O atoms in each oxygen layer.
Li and oxygen atoms were systematically removed from the active regions of the 7~slab models, considering all possible Li and O vacancy decorations within the given model.
Note that for each active region, irrespective of termination, only the surface O atoms were taken into consideration.
The inversion symmetry of the slab models was always maintained when creating Li and O vacancies, i.e., a symmetrically equivalent modification was introduced at the bottom of the slab models.
In total, 1052 surface models were generated using this approach, all of which were used as input for density-functional theory calculations.

\subsection{Density-functional theory calculations}
\label{sec:DFT}

All density-functional theory (DFT)~\cite{pr136-1964-B864, pr140-1965-A1133} calculations were performed using the \textit{Vienna Ab initio Simulation Package} (VASP) software~\cite{prb54-1996-11169, cms6-1996-15} and the projector augmented wave (PAW) approach~\cite{prb50-1994-17953, prb59-1999-1758}.
The electronic wave functions were represented in a plane-wave basis set with an energy cutoff of 520~eV.
We used $\Gamma$-centered $k$-point meshes with $N_{i}= \lfloor{}\max(1, 25 |\vec{b}_{i}|)\rfloor{}$ points in reciprocal direction $i$ for the integration of the Brillouin zone, where $b_i$ is the $i$-th reciprocal lattice vector.
Gaussian smearing with a width of 0.05~eV and an energy convergence criterion of $1\times10^{-5}$ eV were used in the self-consistent field calculations.

Two levels of theory were considered.
For the optimization of slab geometries, the generalized gradient approximation (GGA) exchange-correlation functional by Perdew, Burke and Ernzerhof (PBE)~\cite{prl77-1996-3865} was used.
Additionally, a rotationally invariant Hubbard-U term~\cite{prb57-1998-1505} ($U=6$~eV \cite{cms50-2011-2295}) was employed to correct for the GGA self-interaction error in the description of the strongly correlated Ni \textit{d} electrons.
The convergence thresholds for atomic forces in geometry optimizations was $1\times10^{-4}$ eV \AA{}$^{-1}$.

PBE exhibits a systematic error for oxide formation energies, and the empirical correction of $-1.36$~eV per \ce{O2} by Wang et al.\ was employed to obtain more accurate relative stabilities~\cite{prb73-2006-195107}.
However, the Wang correction is dependent on the oxygen valence state ($-2$ in the bulk oxide), and oxygen loss might involve also the formation of oxidized lattice oxygen species.
For a more detailed discussion see the \textbf{Supporting Section~S1} and \textbf{Figures~S1 and~S2}.

Therefore, all surface phase diagrams reported in the following were determined based on calculations with the Strongly Constrained and Appropriately Normed (SCAN)~\cite{prl115-2015-036402} meta-GGA functional, which does not exhibit any systematic error for oxygen species~\cite{prm6-2022-035003} and has previously been validated for LIB cathode materials~\cite{nc8-2016-831, ncm4-2018-1}.
We performed single point SCAN+rVV10+U ($U=0.6$~eV \cite{prm6-2022-035003}) calculations for structures that were previously optimized using PBE+U with dispersive van-der-Waals interactions considered by including the revised Van-Voorhis approach~\cite{prx6-2016-041005, prl103-2009-063004, tjocp133-2010-244103}, which was shown to further improve the accuracy of formation energies~\cite{prm2-2018-063801, prm6-2022-035003}.

The valence state of Ni atoms was determined based on the magnetic moment, which in turn was calculated by integration of the spin density difference within a sphere around the ionic center~\cite{esl5-2002-A145}.
Changes in the valence states of oxygen atoms were further determined using Bader charge analysis~\cite{cms36-2006-354, jcc28-2007-899, jpcm21-2009-084204, tjocp134-2011-064111} and by comparison of \ce{Ni-O} bond lengths in optimized structures.
The Python Materials Genomics (pymatgen) toolkit was used to generate all input files for DFT calculations~\cite{cms50-2011-2295, cms68-2013-314}.

\subsection{First-principles surface phase diagrams}
\label{sec:surface-phase-diagram-approach}

In thermodynamic equilibrium, the \ce{LiNiO2} surface forms the reconstruction with the lowest Gibbs free energy for the given conditions.
The grand-canonical surface phase diagram is thus determined by the surface free energies of all possible surface structures with different arrangements of lithium and oxygen vacancies, subject to the chemical potentials of lithium and oxygen.

To derive an expression of the surface free energy, we consider the formal truncation of the extended \ce{LiNiO2} crystal structure along a lattice plane.
In thermodynamic equilibrium with oxygen and lithium reservoirs (e.g., the reactants during the synthesis of the material), the surfaces may absorb or release Li and O atoms, so that the surface stoichiometry can differ from the stoichiometry of the \ce{LiNiO2} bulk.
A surface slab model of any surface, whether ideal stoichiometric or reconstructed, can thus be thought of as the result of the formal formation reaction
\begin{widetext}
\begin{align}
\ce{
 $\frac{n_{\ce{Ni}}^{\textup{slab}}}{n_{\ce{Ni}}^{\textup{bulk}}}$
Li_{$n_{\ce{Li}}^{\textup{bulk}}$}Ni_{$n_{\ce{Ni}}^{\textup{bulk}}$}O_{$n_{\ce{O}}^{\textup{bulk}}$}
  + $\Bigl( n_{\ce{Li}}^{\textup{slab}} -
    \frac{n_{\ce{Ni}}^{\textup{slab}}}{n_{\ce{Ni}}^{\textup{bulk}}}
      n_{\ce{Li}}^{\textup{bulk}} \Bigr)$
  Li
  + \frac{1}{2}$\Bigl( n_{\ce{O}}^{\textup{slab}} -
    \frac{n_{\ce{Ni}}^{\textup{slab}}}{n_{\ce{Ni}}^{\textup{bulk}}}
      n_{\ce{O}}^{\textup{bulk}}\Bigr)$
  O2
  ->
  Li_{$n_{\ce{Li}}^{\textup{slab}}$}Ni_{$n_{\ce{Ni}}^{\textup{slab}}$}O_{$n_{\ce{O}}^{\textup{slab}}$}
  }
  \quad .
  \label{eq:slab-formation-reaction}
\end{align}
\end{widetext}
The \emph{surface free energy} is the reaction free energy of reaction~\eqref{eq:slab-formation-reaction} normalized by the surface area $A$ and is given by
\begin{align}
\begin{aligned}
  \gamma = \frac{1}{2A}\Bigl[
  G_{\textup{slab}} &- \frac{n_{\ce{Ni}}^{\textup{slab}}}{n_{\ce{Ni}}^{\textup{bulk}}} G_{\textup{bulk}}
  \\
	&- \sum_{\textup{i}}^{\ce{Li},
  \textup{O}}(n_{\textup{i}}^{\textup{slab}} -
  \frac{n_{\textup{Ni}}^{\textup{slab}}}{n_{\ce{Ni}}^{\textup{bulk}}}
  n_{\textup{i}}^{\textup{bulk}}) \mu_{\textup{i}}
  \Bigr]
  \quad ,
\end{aligned}
  \label{eq:surface-energy}
\end{align}
where $n_{i}^{\textup{slab}}$ and $n_{i}^{\textup{bulk}}$ are the number of atoms of species $i$ (\ce{O} and \ce{Li}) in the slab and bulk models, respectively, $G_{\textup{slab}} = G(\ce{Li_{$n_{\ce{Li}}^{\textup{slab}}$}Ni_{$n_{\ce{Ni}}^{\textup{slab}}$}O_{$n_{\ce{O}}^{\textup{slab}}$}})$ and $G_{\textup{bulk}} = G(\ce{Li_{$n_{\ce{Li}}^{\textup{bulk}}$}Ni_{$n_{\ce{Ni}}^{\textup{bulk}}$}O_{$n_{\ce{O}}^{\textup{bulk}}$}})$ are the Gibbs free energy of the slab and bulk models, respectively, and the \ce{Ni} content is assumed to be constant.
For the fully lithiated \ce{LiNiO2} bulk composition, the number of Li and Ni atoms is identical, and we identify $n_{\ce{Li}}^{\textup{bulk}} = n_{\ce{Ni}}^{\textup{bulk}}$, and $n_{\ce{O}}^{\textup{bulk}} = 2n_{\ce{Ni}}^{\textup{bulk}}$.
Neglecting the temperature dependence of the solids, $G_{\textup{slab}}$ and $G_{\textup{bulk}}$ can be obtained from DFT calculations, representing the energies of the slab model and the bulk structure (one \ce{LiNiO2} formula unit).

In equilibrium, the chemical potential of Li, $\mu_{\textup{Li}}$, is equal to the sum of the Li ion and electron chemical potentials, $\mu_{\textup{Li}} = \mu_{\ce{Li+}} + \mu_{\ce{e-}}$.
Using the Li metal electrode \ce{Li <=> Li+ + e-} as reference, we introduce the reference chemical potentials $\mu_{\ce{Li+}}^{\circ}$ and $\mu_{\ce{e-}}^{\circ}$ with $\mu_{\ce{Li+}}^{\circ} + \mu_{\ce{e-}}^{\circ}
  = G\bigl(\ce{Li}_{\textsc{bcc}}\bigr)$, where $G\bigl(\ce{Li}_{\textsc{bcc}}\bigr)$ is the free energy of Li metal in the body-centered cubic structure, which is also approximated with the zero-Kelvin DFT energy.
The difference of the \ce{Li+} and \ce{e-} chemical potentials from the reference Li metal electrode determines the cell potential $V$: $\Delta\mu_{\ce{Li+}} + \Delta\mu_{\ce{e-}}
  = - F\, V$, where $F$ is Faraday's constant.

The chemical potential of oxygen depends, in principle, on the temperature and pressure, though the pressure dependence is negligible compared to the temperature dependence~\cite{jmca-2020-10.1039.D0TA07706B}.
Ignoring the impact of pressure, the oxygen chemical potential can be expressed as~\cite{prl96-2006-107203}
\begin{align}
\begin{aligned}
	\mu_{\textup{O}}(T)
	= \frac{1}{2}\Bigl\{
	\mu_{\ce{O2}}^{0\textup{K}}
    &+ \bigl[ \Delta H^{\circ} + \Delta H(T) \bigr] \\
	&- T\bigl[ S^{\circ} + \Delta S(T) \bigr]
	\Bigr\}
\quad ,
\end{aligned}
   \label{eq:O-chemical-potential}
\end{align}
where $\frac{1}{2}$ is a normalization factor accounting for the two oxygen atoms in each \ce{O2} molecule, and $\mu_{\textup{\ce{O2}}}^{0\textup{K}}$ is the oxygen chemical potential at 0~Kelvin (equal to the enthalpy), which was obtained from DFT calculations as described above.
The remaining terms are the enthalpy and entropy contributions to the relative oxygen chemical potential:
The difference of the enthalpy at standard conditions from its 0~Kelvin value, $\Delta{}H^{\circ}$, and the standard entropy $S^{\circ}$ were taken from the NIST-JANAF Thermochemical Tables~\cite{NIST-JANAF-1998}.
$\Delta H(T) = C_p(T - T_{\textup{0}})$ and $\Delta S(T) = C_p \ln{(T - T_{\textup{0}})}$, where the heat capacity $C_{p} = 3.5k_B$ was taken to be the value for an ideal gas of diatomic molecules at $T \geq 298$~K~\cite{prl96-2006-107203}.

Finally, taken together, the surface energy of any \ce{LiNiO2} surface (ideal stoichiometric or defected/reconstructed) can be approximated as:
\begin{widetext}
\begin{align}
	\gamma(T, V)
  	\approx \frac{1}{2A} \Bigl\{
    E_{\textup{slab}}
    + (n_{\ce{Ni}} - n_{\ce{Li}}) \Bigl[
    E\bigl(\ce{Li}_{\textsc{bcc}}\bigr)
    -  F\, V
    \Bigr]
    + \frac{1}{2} (2n_{\ce{Ni}} - n_{\ce{O}}) \mu_{\ce{O2}}(T)
   	- n_{\ce{Ni}} E\bigl(\ce{LiNiO2}\bigr)
  \Bigr\}
\label{eq:final-equation}
\quad ,
\end{align}
\end{widetext}
where $E$ denotes DFT energies, and $E_{\textup{slab}} = E(\ce{Li_{$n_{\ce{Li}}^{\textup{slab}}$}Ni_{$n_{\ce{Ni}}^{\textup{slab}}$}O_{$n_{\ce{O}}^{\textup{slab}}$}})$.

Equation~\eqref{eq:final-equation} describes free-energy planes as function of the temperature $T$ (via the oxygen chemical potential) and cell potential $V$ (via the lithium chemical potential).
The surface phase diagram is the two-dimensional projection of the three-dimensional free energy planes in the direction of the free energy axis (see also \textbf{Supporting Figure~S3}).

Equation~\eqref{eq:final-equation} is general and can describe both Li- and Ni-terminated slab models.
However, the free energy of the Li-terminated \ce{LiNiO2(001)} slab (\textbf{Figure~\ref{fig:active-regions}a}) after removal of the active Li atoms and the outermost O atoms (\ce{Li_{8}Ni_{12}O_{24}}) is slightly different from the free energy of the Ni-terminated slab (\ce{Li_{12}Ni_{16}O_{32}}, \textbf{Figure~\ref{fig:active-regions}d}) because the DFT energy exhibits a small dependence on the number of layers in the slab model.
We compensated for this unwanted discrepancy by applying a constant shift corresponding to the energy difference of the two slab models, $\Delta \gamma = \gamma_{\textup{Li-term}} - \gamma_{\textup{Ni-term}} = \frac{1}{2A} \bigl\{E(\ce{Li_{8}Ni_{12}O_{24}}) - E(\ce{Li_{12}Ni_{16}O_{32}}) + 4E(\ce{LiNiO2})\bigr\} = 0.001$~eV/\AA{}$^2$, to the energy of the Ni-terminated structure models, so that energies from both slab models could be combined to construct the surface phase diagram of the $(001)$ facet.

\section{Results}
\label{sec:results}

\begin{figure*}[tp]
  \centering
  \includegraphics[width=0.9\textwidth]{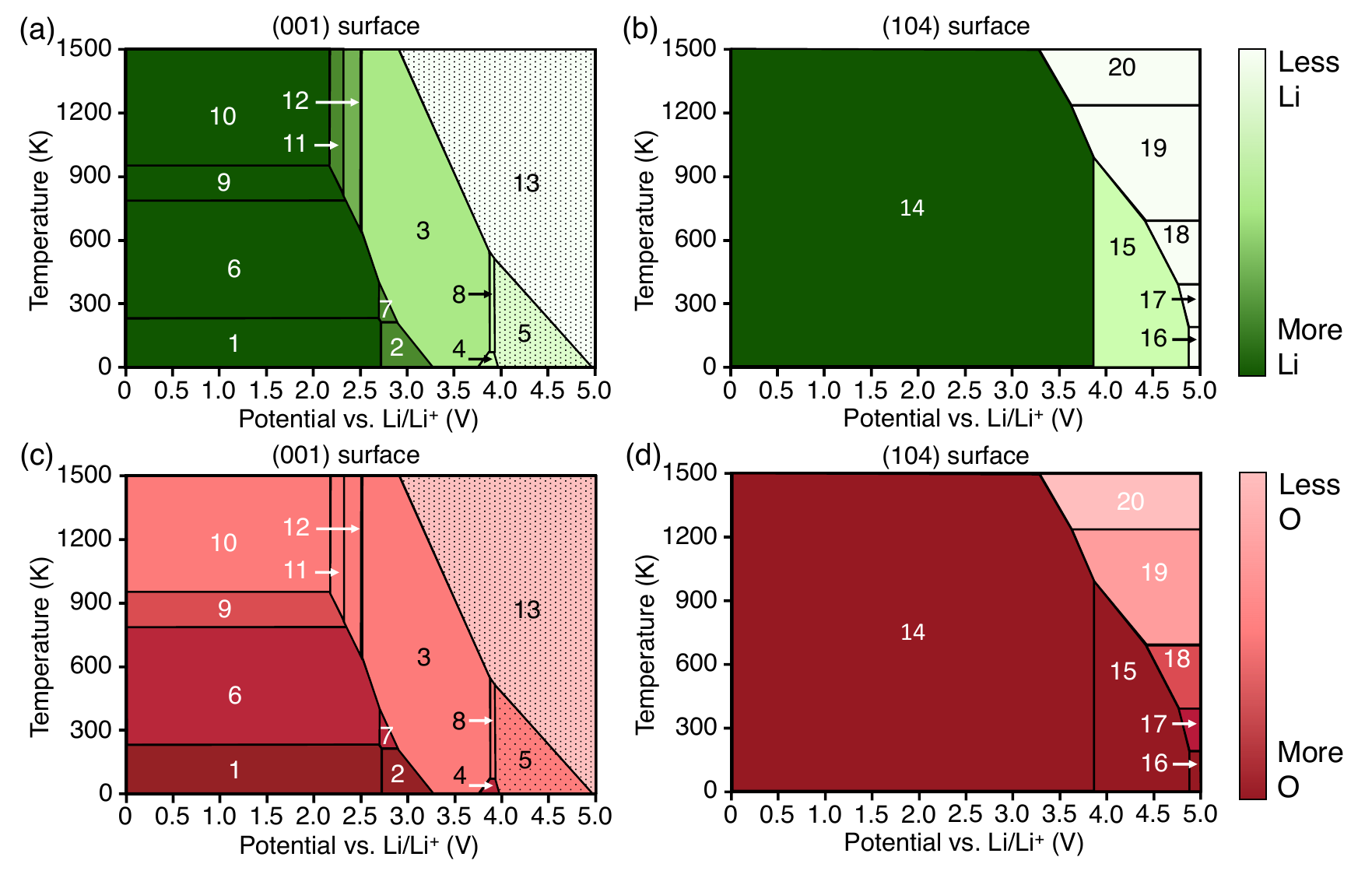}
  \caption{\label{fig:equilibrium-PDs}%
    Equilibrium surface phase diagrams of the \ce{LiNiO2(001)} and \ce{(104)} surfaces.  The diagrams show the predicted thermodynamic stability regions of surface reconstructions as a function of the open-circuit potential and the temperature.  The color gradient indicates the lithium content in \textbf{(a)}~and~\textbf{(b)} (shades of green) and the oxygen content in \textbf{(c)}~and~\textbf{(d)} (shades of red).  Dotted fill patterns indicate surface reconstructions based on Ni-terminated surface models from \textbf{Figure~\ref{fig:active-regions}}.  More details of the various surface phases are given in \textbf{Table~\ref{tbl:phase-compositions}} and \textbf{Supporting Table~S1}.}
\end{figure*}

\begin{table*}[tbp]
  \centering
  \caption{\label{tbl:phase-compositions}%
    Percentage of lithium and oxygen in the active regions of the surface slab models for the phases occurring in the phase diagrams of \textbf{Figure~\ref{fig:equilibrium-PDs} and~\ref{fig:cycling-PDs}}.}
    \vspace{0.5\baselineskip}
  \footnotesize\hspace*{-4mm}
  \begin{tabular}{l*{14}{c}}
    \toprule
    & \multicolumn{13}{c}{\textbf{(001) Surface Phases}} \\
    \cmidrule{2-15}
    & \textbf{1} & \textbf{2} & \textbf{3} & \textbf{4} & \textbf{5} & \textbf{6} & \textbf{7} & \textbf{8} & \textbf{9} & \textbf{10} & \textbf{11} & \textbf{12} & \textbf{13} & \textbf{13$^{*}$}\\
    \midrule
    Surface Li (\%) & 100 & 87.5 & 62.5 & 50.0 & 25.0 & 100  & 87.5 & 50.0 & 100  & 100  & 87.5 & 75.0 & 0.0 & 50.0 \\
    Surface O (\%)  & 100 & 100  & 50.0 & 75.0 & 50.0 & 75.0 & 75.0 & 50.0 & 62.5 & 50.0 & 50.0 & 50.0 & 0.0 & 0.0  \\
    \midrule
    & \multicolumn{13}{c}{\textbf{(104) Surface Phases}} \\
    \cmidrule{2-15}
    & \textbf{14} & \textbf{15} & \textbf{16} & \textbf{17} & \textbf{18} & \textbf{19} & \textbf{20} & \textbf{21} & \textbf{22} & \textbf{23} \\
    \midrule
    Surface Li (\%) & 100 & 50.0 & 0.0 & 0.0  & 0.0  & 0.0  & 0.0 & 25.0 & 50.0 & 100.0 \\
    Surface O (\%)  & 100 & 100  & 100 & 75.0 & 50.0 & 25.0 & 0.0 & 75.0 & 75.0 & 75.0  \\
    \bottomrule
  \end{tabular}
\end{table*}

\subsection{Equilibrium surface phase diagrams of the (001) and (104) surfaces}
\label{sec:equilibrium-PDs}

\textbf{Figure~\ref{fig:equilibrium-PDs}} shows the equilibrium surface phase diagrams of the \ce{LiNiO2(001)} and \ce{(104)} surfaces that were constructed considering all enumerated Li/vacancy and O/vacancy orderings (section~\ref{sec:reconstruction-models}).
As seen in the phase diagrams, the lithium content is predicted to decrease with increasing potential (\textbf{Figure~\ref{fig:equilibrium-PDs}a,b}), and the oxygen content decreases with increasing temperature (\textbf{Figure~\ref{fig:equilibrium-PDs}c,d}), in agreement with our expectations.
The lithium and oxygen contents of the various surface phases are listed in \textbf{Table~\ref{tbl:phase-compositions}}, and the information needed to construct the free-energy planes of all considered surface reconstructions is given in \textbf{Supporting Table~S1}.

Apart from the general trends with Li and O content, the phase diagrams of the two surfaces differ significantly.
The phase diagram of the $(104)$ surface (\textbf{Figure~\ref{fig:equilibrium-PDs}b,d}) is overall simpler and exhibits fewer phases.
For a wide potential ($V<3.8$~V) and temperature ($T<1\,000$~K) range, the $(104)$ surface is predicted to remain fully lithiated without releasing oxygen.
At temperatures below $\sim{}1\,000$~K, the $(104)$ surface undergoes two phase transitions as Li is extracted.
First, when the potential exceeds $\sim{}3.8$~V, the surface active region is delithiated by 50\% without any oxygen release.
The surface is then fully delithiated at a potential that decreases with the temperature from $>4.8$~V at room temperature.
Above $\sim{}1\,000$~K, the equilibrium potential for complete delithiation is predicted to be so low ($<3.8$~V) that the stability region of the half-delithiated surface phase disappears, i.e., the surface is either fully lithiated or fully delithiated.
The oxygen content in the fully delithiated surface region also decreases with the temperature.

In contrast to the $(104)$ surface, the $(001)$ surface releases oxygen with increasing temperature even when it is fully lithiated (\textbf{Figure~\ref{fig:equilibrium-PDs}a,c}).
The surface also shows a more gradual decrease of the lithium content with increasing potential.
These differences in the phase diagram reflect the change of the surface termination with increasing potential, since the $(001)$ surface is preferentially Li-terminated at potentials below $\sim$3.9~V and Ni-terminated at greater potentials.

\begin{figure*}
  \centering
  \includegraphics[width=0.8\textwidth]{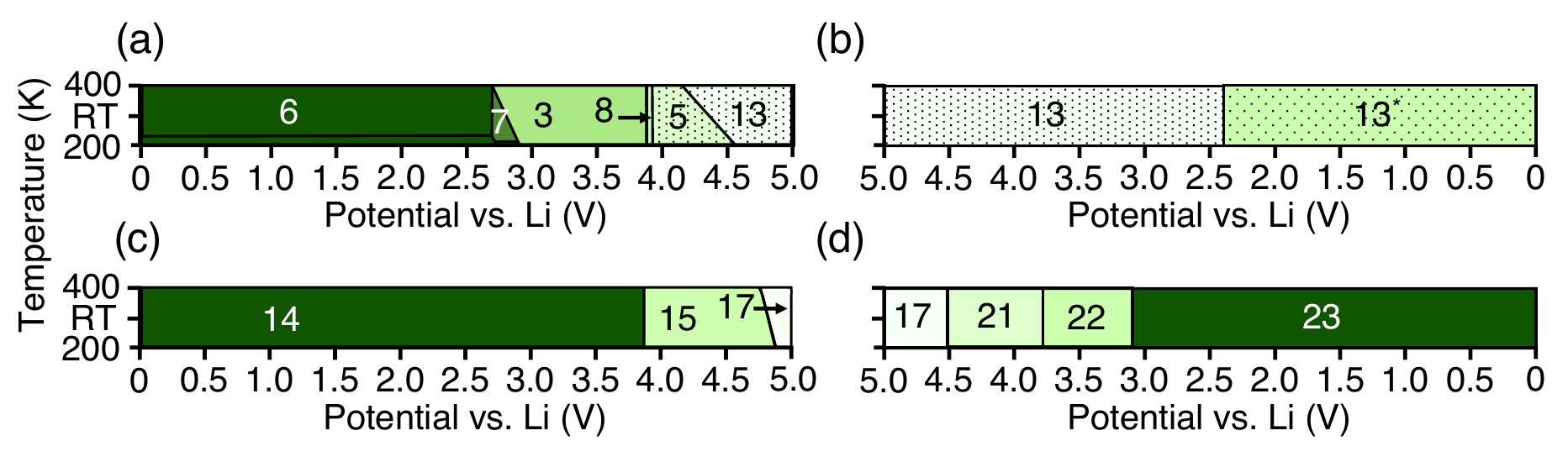}
  \caption{\label{fig:cycling-PDs}%
    Predicted surface phase diagrams of the \textbf{(a,b)}~\ce{LiNiO2(001)} and \textbf{(c,d)}~\ce{LiNiO2(104)} surfaces during cycling near room temperature.  Oxygen release was assumed to be irreversible.  The discharge phase diagrams \textbf{(b) and (d)} assume that the surface was fully delithiated during charge.  Partial delithiation and the impact of the voltage cutoff and temperature are discussed in the following section.}
\end{figure*}

\subsection{Surface phases visited during electrochemical cycling}
\label{sec:charge-discharge}

During the initial charge of a battery with \ce{LiNiO2} cathode, the surface phase diagrams of \textbf{Figure~\ref{fig:equilibrium-PDs}} are, in principle, traversed from the left (fully lithiated) to the right (fully delithiated).
However, the phase diagrams of \textbf{Figure~\ref{fig:equilibrium-PDs}} correspond to thermodynamic equilibrium conditions, whereas during battery cycling reversibility can only be expected for the lithium content, not for the oxygen content.
\ce{O2} gas released during cycling has to be considered lost, and therefore the oxygen content in the surface phases visited during charge and discharge can only decrease and never increase.
This also implies that the surface phases visited during discharge (re-lithiation) have to be different if oxygen was lost during charge.

\textbf{Figure~\ref{fig:cycling-PDs}} shows non-equilibrium surface phase diagrams that account for the irreversible nature of oxygen release.
The shown discharge phase diagrams were constructed assuming full delithiation of the surface regions during charge, and diagrams for partial delithiation (corresponding to cutoff potentials below 5~V) are shown in \textbf{Supporting Figure~S4}.

As seen in \textbf{Figure~\ref{fig:cycling-PDs}b}, the $(001)$ surface exhibits a simple discharge phase diagram at room temperature, visiting only two phases.
Note that the surface does not fully relithiate, since oxygen release during charge resulted in the loss of lithium sites.
In contrast, the stoichiometric $(104)$ surface releases less oxygen during charge, 25\% of the oxygen in the surface layers, and complete relithiation is predicted during discharge.
Interestingly, our calculations predict a gradual relithiation of the $(104)$ surface.
We also note that during charge, lithium extraction from the $(001)$ surface is predicted to begin nearly 1~V below the potential at which the first lithium is extracted from the $(104)$ surface.

\begin{figure*}[tbp]
  \centering
  \includegraphics[width=1.0\textwidth]{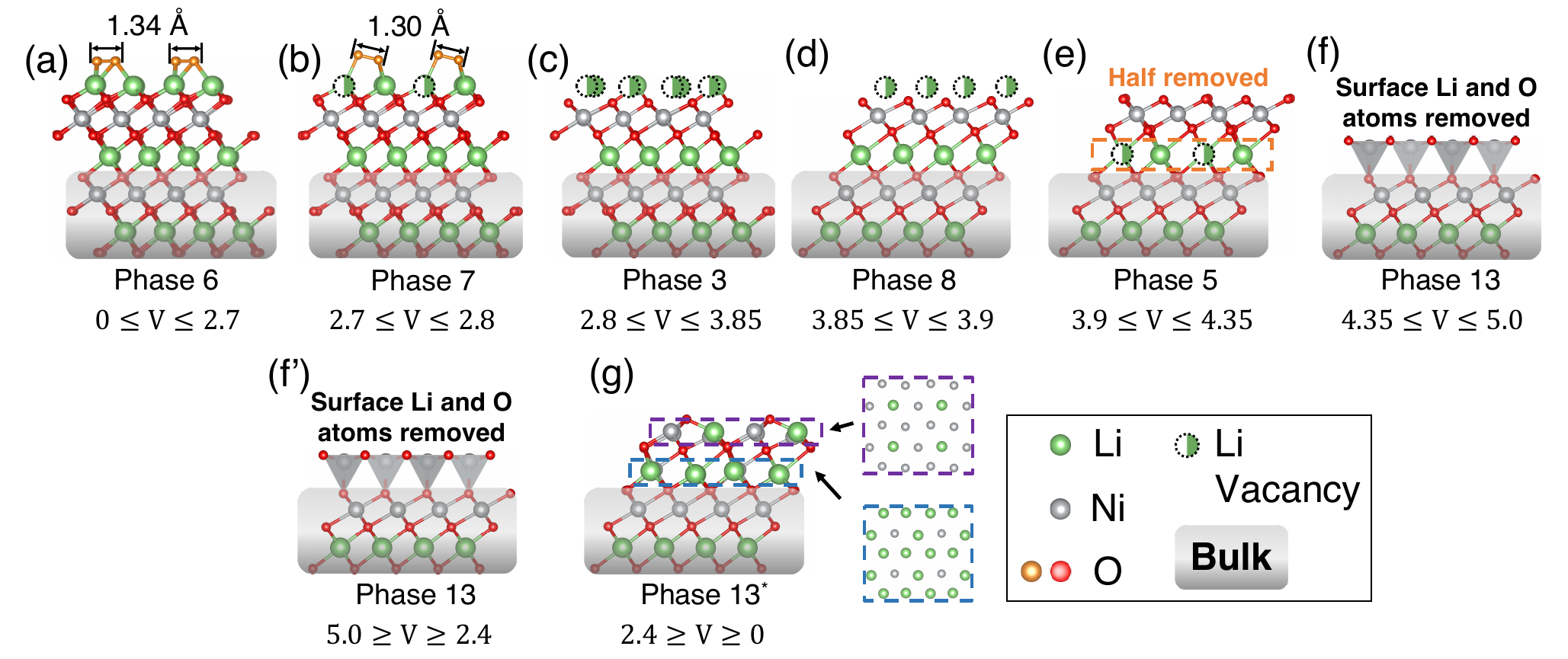}
  \caption{\label{fig:001-degradation-pathways}%
    Reconstructions of the \ce{LiNiO2(001)} surface corresponding to the surface phases visited during \textbf{(a)--(f)}~charge and \textbf{(f')--(g)}~discharge at room temperature.  The path through the phase diagram is shown in \textbf{Figure~\ref{fig:cycling-PDs}}.  The Li content gradually decreases with increasing potential.  \textbf{(f)}~At the end of charge, Ni atoms have migrated into tetrahedral sites within the Li layer beneath.  \textbf{(g)}~Relithiation leads to the formation of a cation-mixed (spinel-like) surface phase.  The Ni/Li ordering is shown in the insets of subfigure~(g).  Adsorbed oxygen is colored gold, whereas lattice oxygen is red.}
\end{figure*}

\begin{figure*}[tbp]
  \centering
  \includegraphics[width=0.9\textwidth]{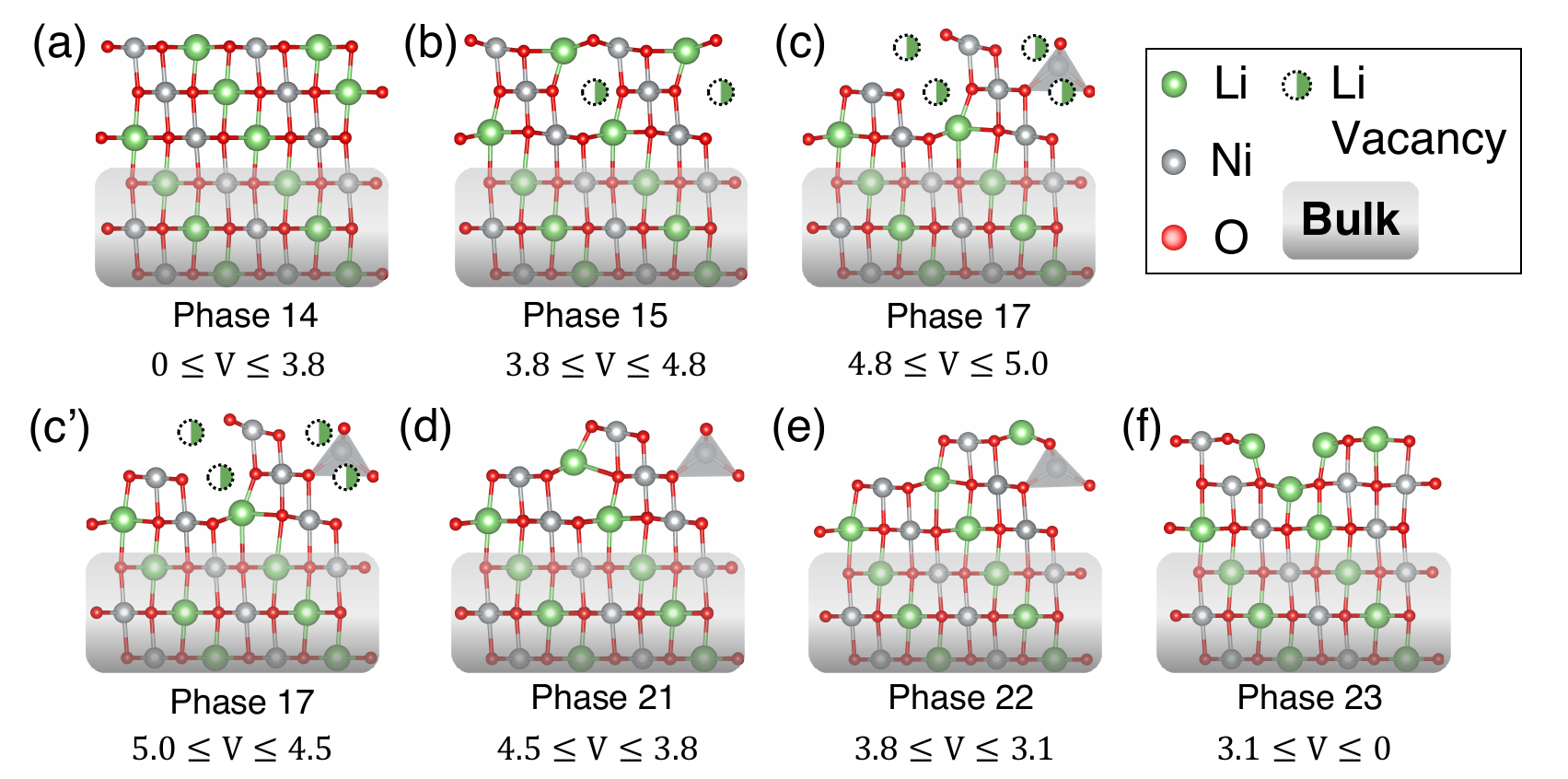}
  \caption{\label{fig:104-degradation-pathways}%
    Reconstructions of the \ce{LiNiO2(104)} surface visited during the first \textbf{(a)--(c)}~charge and \textbf{(c')--(f)}~discharge cycle at room temperature.  The corresponding phase diagrams are shown in \textbf{Figure~\ref{fig:cycling-PDs}c,d}.  Upon relithiation, Ni/Li mixed phases (phases 17, 21, 22, and 23) are predicted to be thermodynamically stable.  Colors are as in \textbf{Figure~\ref{fig:001-degradation-pathways}}.}
\end{figure*}

The atomic structures of the surface reconstructions that are visited during charge and discharge are depicted in \textbf{Figure~\ref{fig:001-degradation-pathways}} and \textbf{Figure~\ref{fig:104-degradation-pathways}} for the $(001)$ and $(104)$ surfaces, respectively.
As seen in \textbf{Figure~\ref{fig:001-degradation-pathways}a}, in its discharged state, the $(001)$ surface is Li terminated, and additional \ce{O2} molecules are adsorbed on the surface.
Based on the \ce{O-O} bond length of 1.34~\AA{} and Bader charge analysis (see \textbf{Supporting Table~S2}), the adsorbed oxygen corresponds to reduced peroxide or superoxide species, indicating an already reactive state.
In an electrochemical cell, such activated oxygen species are unlikely to be present and would react with the electrolyte or binder, however, the predicted surface phase reflects the general reducing conditions for which phase~6 is predicted to be stable.
As the potential increases (\textbf{Figure~\ref{fig:001-degradation-pathways}b}), the Li contents in the surface layers decreases and the adsorbed oxygen species are further oxidized, indicated by the contraction of the \ce{O-O} bond length (\textbf{Supporting Table~S2}).
At a potential of around 2.8~V, all surface oxygen atoms are fully oxidized and released as neutral \ce{O2} gas, leaving undercoordinated Li atoms on the surface (\textbf{Figure~\ref{fig:001-degradation-pathways}c}).
The remaining surface Li becomes unstable at a potential of $\sim$3.85~V (phase~8), and sub-surface layers also begin to be delithiated (\textbf{Figure~\ref{fig:001-degradation-pathways}d--e}).
Finally, at an equilibrium potential of around 4.35~V, the surface region is fully delithiated, the surface Ni atoms migrate into tetrahedral sites within the empty Li layer, and the coordinating oxygen atoms are oxidized and lost   (\textbf{Figure~\ref{fig:001-degradation-pathways}f}).

Owing to the oxygen release and Ni migration to Li vacancies, Li sites are lost, and the discharge capacity of the surface layers is lower than the charge capacity.
Upon relithiation, the \ce{LiNiO2}(001) surface undergoes a reconstruction from \textbf{Figure~\ref{fig:001-degradation-pathways}f'} to \textbf{Figure~\ref{fig:001-degradation-pathways}g}, in which only 50~\% surface Li is retained.
The final discharge phase is Ni rich and cation mixed with spinel-like Ni/Li ordering (\textbf{Figure~\ref{fig:001-degradation-pathways}g}).

\textbf{Figure~\ref{fig:104-degradation-pathways}} shows, the surface phases of the \ce{LiNiO2(104)} surface that are visited during the first charge-discharge cycle assuming full delithiation upon charge.
At room temperature, the oxygen content within the stoichiometric $(104)$ surface varies less than in the $(001)$ surface, and the structure of the surface region remains mostly intact during delithiation (\textbf{Figure~\ref{fig:104-degradation-pathways}a--c}).
As the potential reaches 3.8~V, half of the surface Li atoms are extracted, and Li vacancies are formed below the surface, whereas the Li sites in the topmost layer remain occupied.
Although Li atoms can only be extracted from exposed surface sites, sub-surface Li vacancies can form when the remaining Li rearranges to the thermodynamically preferred atomic ordering.
No oxygen release is predicted before all of the surface Li is extracted at a potential of around 4.8~V, and even then only a quarter of the oxygen atoms in the immediate surface layer become unstable.
However, as for the $(001)$ surface, Ni atoms are predicted to migrate to tetrahedral sites in the delithiated Li layer when the surface region is fully delithiated (\textbf{Figure~\ref{fig:001-degradation-pathways}f}).
Upon discharge, the tetrahedral Ni remains stable in the Li layer even when half of the original Li content is reintercalated into the \ce{LiNiO2(104)} surface (\textbf{Figure~\ref{fig:104-degradation-pathways}e}).
Further Li insertion at potentials below 3.1~V triggers additional Ni migration to the subsurface, so that at the end of the first discharge a spinel-like Ni/Li ordering is thermodynamically preferred (\textbf{Figure~\ref{fig:104-degradation-pathways}f}).

\subsection{Impact of temperature and voltage range}
\label{sec:T-and-V-range}

\begin{figure*}
  \centering
  \includegraphics[width=0.9\textwidth]{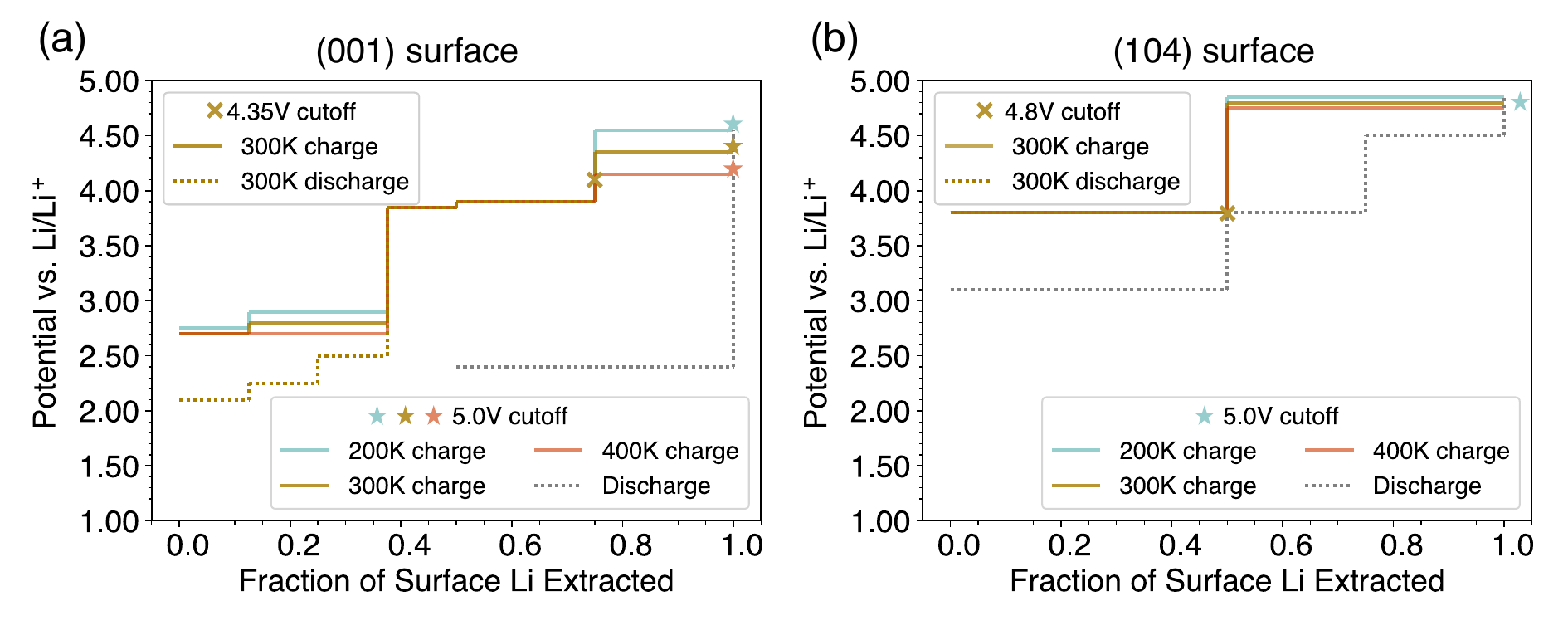}
  \caption{\label{fig:T-and-V-dependence}%
    Voltage profiles for different temperature and potential ranges.  \textbf{(a)}~Voltage profiles of the \ce{LiNiO2(001)} surface at 200~K, 300~K, and 400~K and with potential cutoffs of 4.35~V and 5.0~V, respectively.  For the 5.0~V cutoff, the discharge profile is temperature independent.  \textbf{(b)}~Equivalent voltage profiles of the \ce{LiNiO2(104)} surface at 200~K, 300~K, and 400~K but with potential cutoffs of 4.8~V and 5.0~V, respectively.}
\end{figure*}

In the previous sections, we only considered cycling near room temperature and with complete surface delithiation during the first charge.
However, the degree of delithiation during the first charge, and thus the phases visited during cycling, depend on the temperature and potential range.

\textbf{Figure~\ref{fig:T-and-V-dependence}} shows how the charge-discharge voltage profiles of the surface region vary with the temperature (200~K~$\leq{}T\leq$~400~K) and the upper voltage limit ($4.35$~V and $5.0$~V for the \ce{LiNiO2(001)} surface, $4.8$~V and $5.0$~V for the \ce{(104)} surface).
As seen in the figure, the voltage profile of the $(001)$ surface is rather sensitive with respect to these variables, and fully reversible capacity can be achieved by reducing the potential cutoff.
Meanwhile, the optimal cutoff also depends on the temperature and would be $4.35$~V at room temperature but only 4.15~V at 100~K higher.

Compared with the $(001)$ surface, the voltage profile of the $(104)$ surface is less sensitive with respect to the temperature and potential ranges.
However, if the voltage cutoff is chosen such that the $(001)$ surface can be fully reinserted, only half of the capacity of the $(104)$ surface would be utilized.

\section{Discussion}
\label{sec:discussion}

\begin{figure}
  \centering
  \includegraphics[width=0.9\linewidth]{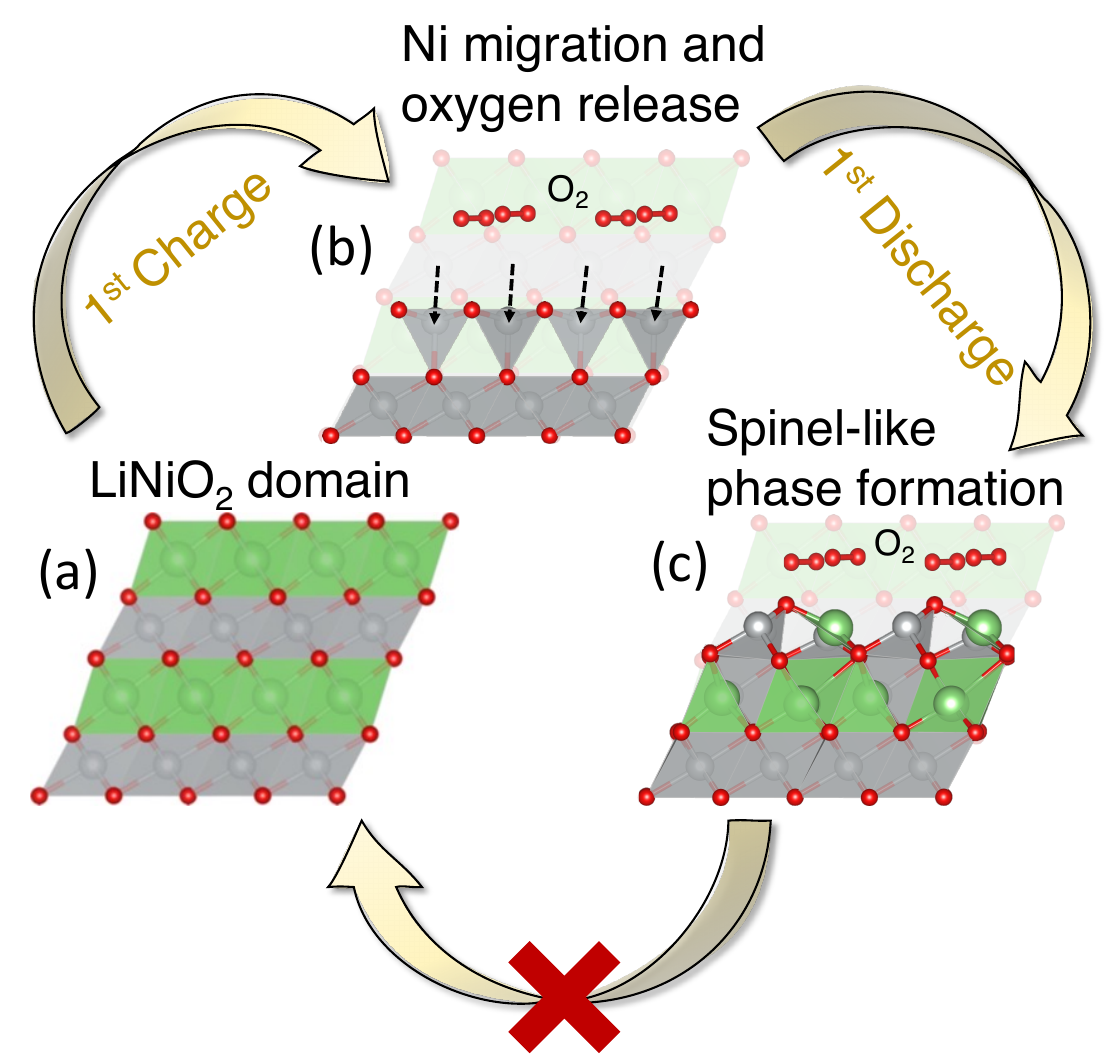}
  \caption{\label{fig:mechanism}%
Schematic of the predicted surface degradation mechanism of \ce{LiNiO2}. \textbf{(a)}~Pristine layered \ce{LiNiO2} before the first charge.  \textbf{(b)}~Oxygen release and Ni migration within the surface regions occurs at the end of charge.  \textbf{(c)}~Oxygen loss leads to the formation of irreversible surface reconstructions.}
\end{figure}

We developed a method for the calculation of intercalation battery electrode surface phase diagrams during electrochemical cycling and determined the phase diagrams of the two most stable \ce{LiNiO2} surfaces, \ce{(001)} and \ce{(104)}, during the first charge-discharge cycle at room temperature.
Our calculations indicate that both surfaces release oxygen during the initial charge, though the polar \ce{LiNiO2(001)} surface releases more oxygen as it undergoes a transition from Li-terminated at low voltages to Ni-terminated at high voltages.
Upon complete delithiation of the surface region of both surfaces, our calculations predict Ni atoms to migrate to tetrahedral sites in the vacated Li layers.
During discharge (relithiation) both $(001)$ and $(104)$ surfaces become Li/Ni mixed and remain permanently altered.
The mechanism is summarized in \textbf{Figure~\ref{fig:mechanism}}.

The finding that the \ce{LiNiO2(001)} surface of the pristine material is Li-terminated is in agreement with a previous computational study~\cite{aami9-2017-33257} that also concluded that the preference for Li termination likely promotes oxygen release at the beginning of charge.

The redox mechanism at the surface of \ce{LiNiO2} involves the participation of both nickel and oxygen.
Our results indicate that Ni and O redox are not sequential but are instead correlated.
At the end of the first charge, when the surface regions are fully delithiated and oxygen has been released, surface reconstructions with tetrahedral Ni species are predicted to be stable for both the $(001)$ and the $(104)$ surface.
Perhaps counter-intuitively, spin integration shows that the tetrahedral Ni is in the $2+$ valence state, i.e., Ni is formally reduced compared to the $3+$ valence state in the pristine material (see \textbf{Supporting Table~S2} and \textbf{Figure~S5}).
This prediction is consistent with the common assumption that \ce{Ni^{4+}} is not stable in tetrahedral sites~\cite{cm18-2006-4768}.
In addition, it has previously been proposed that \ce{Ni^{2+}} ions in the subsurface and surface layers can be stabilized via a strong 180$^{\circ{}}$ superexchange interaction~\cite{afm31-2021-2010291}.
This superexchange interaction originates from $\sigma$ bonds formed between Ni and O states, so that the bonding is significantly covalent and oxidation states cannot be uniquely assigned to the individual Ni and O atoms within a \ce{NiO4} tetrahedron.

Hence, oxygen release from \ce{LiNiO2} and derived Ni-rich cathodes is aggravated by (i)~the ability of Ni to reduce below the 3+ valence state (allowing more oxygen atoms to oxidize), (ii)~the preference of \ce{Ni^{2+}} for tetrahedral coordination (requiring less O in the surface region), and (iii)~the relative mobility of the Ni atoms facilitating the migration onto Li vacancies.

The surface reconstruction mechanism predicted by our calculations for \ce{LiNiO2} is also consistent with experimental observations for Ni-rich NMCs.
In NMC111 (\ce{LiNi_{1/3}Mn_{1/3}Co_{1/3}O_2}), Ni, Mn, and Co have the valence states 2+, 4+, and 3+, respectively~\cite{jpcc121-2017-8290}.
In Ni-rich NMCs, such as NMC811 (\ce{LiNi_{0.8}Mn_{0.1}Co_{0.1}O_2}), \ce{Ni^{3+}} is introduced~\cite{jpcc121-2017-22628}.
Previous experimental studies suggested that the intrinsic instability of Ni-rich materials is caused by the tendency of \ce{Ni^{4+}} to reduce all the way to \ce{Ni^{2+}} upon oxygen loss~\cite{jops233-2013-121, aami6-2014-15140, aami8-2016-1297, cm15-2003-4476}, in agreement with the reconstructions found in our computations.
Though we note that in cathode compositions with multiple transition-metal species surface segregation may introduce an additional complexity to surface reconstructions.
For NMCs, it has been found that Ni preferentially segregates to the (012) surface, whereas Co tends to segregate to the (104) surface~\cite{aem6-2016-1502455}.
It has further been proposed that such facet-dependent segregation in NMC can be controlled with strain to stabilize a spinel surface phase~\cite{pccp22-2020-24490}.

The phase diagrams reported above for charge-discharge cycling assume the complete delithiation of the surface regions of the \ce{LiNiO2} primary particles.
The fully delithiated surface phases are predicted to be thermodynamically stable at equilibrium (open-circuit) potentials $>4.8$~V, which is above typical charge cutoff potentials.
However, batteries are charged at constant currents (i.e., galvanostatically), and the Li content in particle surface regions can deviate significantly from equilibrium conditions~\cite{jes159-2012-A1967}.
It is therefore plausible that a thin surface region would be fully delithiated during charge.

We note that the present study determined the thermodynamically stable surface reconstructions as a function of the state of charge.
Since some of the reconstructions require the migration of Ni atoms from the Ni layer to the Li layer, we expect that this process is also subject to a kinetic barrier that was not considered in the present work.
Further investigation, both computational and experimental, will be required to understand the impact of kinetics and the cycling rate on the surface reconstruction of Li-ion battery cathodes.

For both surfaces, our calculations predict Li/Ni mixed surface reconstructions at the end of the first charge, on complete delithiation of the surface layers.
This is a direct result of the instability of undercoordinated Ni atoms at the \ce{LiNiO2} surface.
The analysis of the temperature and potential range of section~\ref{sec:T-and-V-range} indicates that cation mixing can be avoided by reducing the charge voltage cutoff, which would however reduce the capacity of the \ce{LiNiO2(104)} surface by 50\%.

Note that, depending on the synthesis conditions, other surface facets in addition to the (001) and (104) surfaces can also display significant surface areas and may therefore impact the surface degradation of \ce{LiNiO2} and related cathode materials.
 In particular, the (010) and (012)~\cite{aami9-2017-33257, jmca-2020-10.1039.D0TA07706B, jpcc121-2017-8290} facets have previously been characterized (see \textbf{Figure~S6} for schematics of the fully lithiated surfaces).
 The present work demonstrates that the polar (001) and the non-polar (104) facets, despite their entirely different structures, exhibit similar reconstruction mechanisms that lead to the formation of tetrahedral Ni at high voltages, followed by Ni migration upon discharge.
 Even these two most stable surfaces undergo an irreversible surface reconstruction.
 It can thus be expected that other, less stable surfaces also experience similar reconstructions during charge.

Finally, we emphasize that the present work focuses on an idealized model system of \ce{LiNiO2} single-crystal surfaces in air, and the surface free energy of equation~\eqref{eq:final-equation} is additionally subject to approximations and the inherent errors of the employed DFT method.
SCAN+U has previously been shown to provide accurate voltage predictions~\cite{prm2-2018-063801, cm32-2020-3447, ncm4-2018-1} so that we estimate the uncertainty of the voltage to be only $\sim$0.3~V, which could affect phases with small stability regions (such as phase~4 in \textbf{Figure~\ref{fig:equilibrium-PDs}}) but would not qualitatively alter the phase diagrams.
The uncertainties in the temperature of phase boundaries have to be expected to be more significant because our expression only accounted for the temperature-dependence of the free energy of \ce{O2}.
However, most stability regions in the direction of temperature are on the order of $\sim{}300$~K or wider, so that the predicted phase diagrams are robust with respect to such inaccuracies.
The impact of the environment in an electrochemical cell on the surface phase diagrams is more challenging to estimate.
During charge, oxidized oxygen species can react with the coating, binder, or electrolyte in contact with the \ce{LiNiO2} electrode, and during discharge the electrolyte could function as a reservoir for oxygen (via reductive decomposition).
We expect that such side reactions not considered here will become more important when multiple cycles are considered and the system is given more time to equilibrate.
Ultimately, a comparison with experiments will be important to determine the charge rate for which the predictions are valid.


\section{Conclusions}
\label{sec:conclusions}

In conclusion, we devised a first-principles methodology for the prediction of the surface reconstructions of intercalation electrodes during charge-discharge cycling and applied the method to the \ce{LiNiO2(001)} and \ce{(104)} surfaces.
Our calculations predict that both surfaces release oxygen gas during the first charge, so that the structure of the surfaces is permanently changed upon discharge.
At the end of the first charge, both surfaces exhibit Ni atoms in tetrahedral sites that migrate to Li vacancies during discharge, leading to Li/Ni mixed discharged phases.
This mechanism can be understood as an initial step in the formation of a spinel-like cation-mixed surface phase, which has previously been observed in experiments.
We further found the $(001)$ surface to be significantly less stable during cycling than the $(104)$ surface because of a voltage-dependent preference for either Li or Ni termination.
Based on these findings, we argue that oxygen release is aggravated in Ni-rich cathodes by the ability of Ni to be reduced below the 3+ valence state, the preference of \ce{Ni^{2+}} for lower (4-fold) coordination, and the mobility of the Ni atoms.
While surface degradation can be prevented by reducing the (temperature-dependent) upper voltage cutoff during galvanostatic charging at the cost of usable capacity, we speculate that compositional changes (such as in NCA and NMC) destabilize the tetrahedral Ni at the end of charge.

The present work is a first step towards understanding and preventing the initial stages of surface degradation in Ni-rich cathodes and can serve as a starting point for the computational design of modified compositions and coatings.
Important next steps will be to investigate the impact of interactions between the electrolyte, coatings, and the cathode surface as well as kinetic limitations and the impact of the cycling rate.


\section{Acknowledgements}
\label{sec:acknowledgements}

This work was supported by the Alfred P.~Sloan Foundation grant number G-2020-12650.
We acknowledge computing resources from Columbia University's Shared Research Computing Facility project, which is supported by NIH Research Facility Improvement Grant 1G20RR030893-01, and associated funds from the New York State Empire State Development, Division of Science Technology and Innovation (NYSTAR) Contract C090171, both awarded April 15, 2010.
H.G. acknowledges financial support by the U.S. Department of Energy (DOE), Office of Energy Efficiency and Renewable Energy (EERE), Vehicle Technologies Office (VTO), Contract No. DE- SC0012704, Advanced Battery Materials Research program (Tien Duong, Program Manager).
The authors thank Joaquin Rodriguez-Lopez, Zheng Li, Alan West, and Jianzhou Qu for helpful discussions.


\newpage

%

\newpage

\onecolumngrid
\clearpage
\appendix

 \renewcommand{\thesection}{S\arabic{section}}
 \renewcommand{\thefigure}{S\arabic{figure}}
 \renewcommand{\thetable}{S\arabic{table}}
 \setcounter{figure}{0}
 \setcounter{table}{0}
 \setcounter{section}{0}

\section{Supporting information}

\subsection{Comparison of PBE+U and SCAN+rVV10+U phase diagrams}
\label{sec:PBE-vs-SCAN}

\begin{figure}[H]
	\centering
	\includegraphics[width=0.9\textwidth]{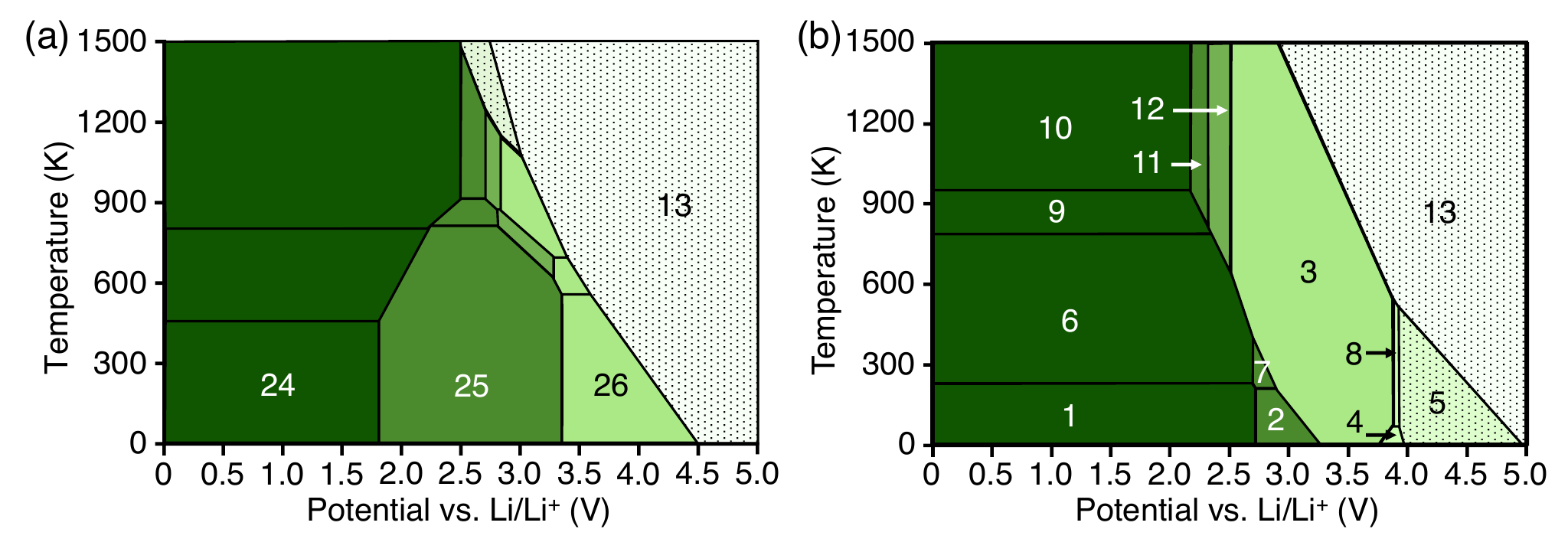}
	\caption{Equilibrium surface phase diagrams of the \ce{LiNiO2(001)} surface as predicted by calculations using \textbf{(a)}~\textbf{PBE+U} with additional oxygen correction and \textbf{(b)}~\textbf{SCAN+rVV10+U}.  The qualitative differences between the PBE+U and SCAN+rVV10+U phase diagram originate from the additional phases 24, 25, and 26 that are artificially stabilized by the PBE+U approach.  The compositions of these extra phases are given in \textbf{Table~\ref{stbl:addtitional-coverage-percent}} and their structures are shown in \textbf{Figure~\ref{fig:001-PBE-U-degradation-pathways}}.}
	\label{fig:reversible-001-pd}
\end{figure}

\begin{figure}[H]
	\centering
	\includegraphics[width=0.8\textwidth]{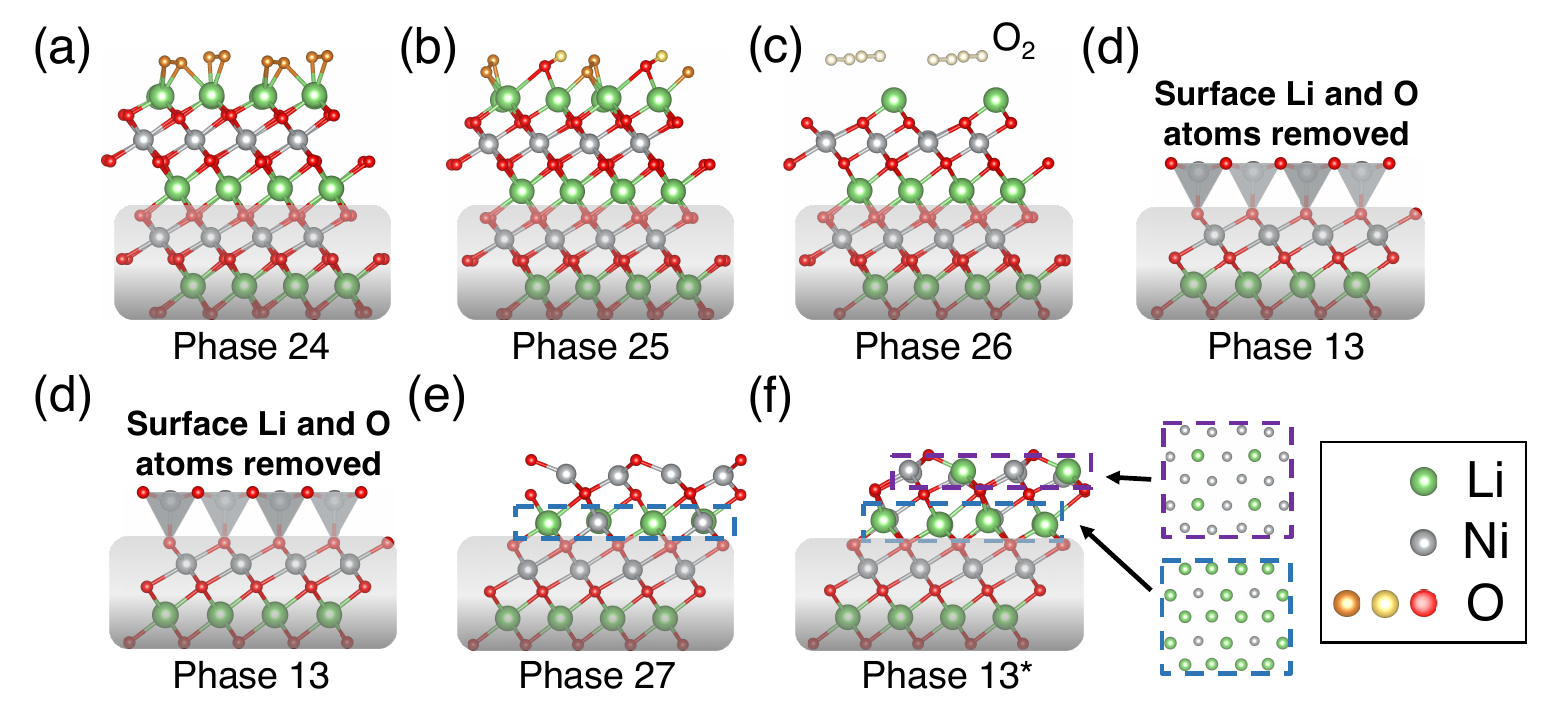}
	\caption{Thermodynamically stable surface reconstructions of the \ce{LiNiO2(001)} surface at room temperature, as predicted by \textbf{PBE+U} calculations.  The corresponding phase diagram is shown in \textbf{Figure~\ref{fig:reversible-001-pd}}.	\textbf{(a-d)}~First charge and \textbf{(d'-f)}~first discharge process.  Surface phases 24, 25, and 26 exhibit adsorbed oxygen with different valence states.  Such structures are artificially overstabilized by the employed PBE+U approach due to inconsistencies in the empirical correction for oxygen overbinding.}
	\label{fig:001-PBE-U-degradation-pathways}
\end{figure}

All surface phase diagrams reported in the main manuscript were constructed using energies from SCAN+rVV10+U calculations, as described in the methods section.
For comparison, \textbf{Figure~\ref{fig:reversible-001-pd}} shows also the surface phase diagram of the \ce{LiNiO2(001)} surface constructed based on PBE+U energies, and significant qualitative differences are seen.
At room temperature, the differences are due to additional phases (numbered here 24, 25, and 26) for which the employed PBE+U approach predicts large stability regions.
However, these phases are artificially stabilized by the empirical oxide correction after Wang~\cite{prb73-2006-195107} employed in our PBE+U calculations to compensate for the oxygen overbinding of the PBE functional.
The value of the correction term depends on the valence state of the oxygen atoms ($-1.36$~eV per \ce{O2} for oxides, i.e., \ce{O2-}), but the surface oxygen in phases 24, 25, and 26 is partially oxidized (\textbf{Figure~\ref{fig:001-PBE-U-degradation-pathways}}), so that no single correction term can be appropriate for all oxygen within the structure.

Hence, the employed PBE+U approach with empirical oxygen correction is not suitable for the prediction of surface phase diagrams involving oxygen release, and therefore only results for SCAN+rVV10+U, which does not exhibit any systematic error for oxygen species~\cite{prm6-2022-035003}, are reported in the main manuscript.

\subsection{Supplementary tables}

\begin{table}[H]
\centering
\caption{Composition and DFT energy of all surface reconstructions appearing in the phase diagrams.  The SCAN+rvv10+U energies of bulk \ce{LiNiO2} and Li metal in the body-centered cubic structure are $E(\ce{LiNiO2})=-36.81$~eV and $E(\ce{Li})=-2.33$~eV, respectively.  Together with equation~(6) in the main manuscript, this information allows constructing the free energy planes of all stable phases and the reconstruction of the surface phase diagrams.\vspace{0.5\baselineskip}}
\begin{tabular}{rccrrrc}
\toprule
\multicolumn{1}{c}{Phase} & \multicolumn{1}{c}{Facet} & Area (\AA{}$^{2}$) & \multicolumn{1}{c}{\# Li} & \multicolumn{1}{c}{\# Ni} & \multicolumn{1}{c}{\# O} & \multicolumn{1}{c}{Energy (eV)} \\
\midrule
1     & (001)  &  28.2  & 16  & 12 & 32 & -507.51     \\
2     & (001)  &  28.2  & 14  & 12 & 32 & -497.53     \\
3     & (001)  &  28.2  & 10  & 12 & 24 & -427.34     \\
4     & (001)  &  28.2  &  8  & 12 & 28 & -439.32     \\
5     & (001)  &  28.2  &  8  & 16 & 32 & -537.81     \\
6     & (001)  &  28.2  & 16  & 12 & 28 & -482.69     \\
7     & (001)  &  28.2  & 14  & 12 & 28 & -472.77     \\
8     & (001)  &  28.2  & 12  & 16 & 32 & -562.41     \\
9     & (001)  &  28.2  & 16  & 12 & 26 & -469.04     \\
10    & (001)  &  28.2  & 16  & 12 & 24 & -454.97     \\
11    & (001)  &  28.2  & 14  & 12 & 24 & -446.08     \\
12    & (001)  &  28.2  & 12  & 12 & 24 & -436.89     \\
13    & (001)  &  28.2  &  4  & 16 & 24 & -461.15     \\
13*   & (001)  &  28.2  & 12  & 16 & 24 & -498.58     \\
14    & (104)  &  31.2  & 18  & 18 & 36 & -659.35     \\
15    & (104)  &  31.2  & 14  & 18 & 36 & -634.93     \\
16    & (104)  &  31.2  & 10  & 18 & 36 & -606.52     \\
17    & (104)  &  31.2  & 10  & 18 & 34 & -594.19     \\
18    & (104)  &  31.2  & 10  & 18 & 32 & -581.43     \\
19    & (104)  &  31.2  & 10  & 18 & 30 & -567.99     \\
20    & (104)  &  31.2  & 10  & 18 & 28 & -553.20     \\
21    & (104)  &  31.2  & 12  & 18 & 34 & -607.68     \\
22    & (104)  &  31.2  & 14  & 18 & 34 & -619.73     \\
23    & (104)  &  31.2  & 18  & 18 & 34 & -641.18     \\
\bottomrule
\label{stbl:free-energy-plane-expressions}
\end{tabular}
\end{table}

\begin{table}[H]
\centering
\caption{Average bond length of adsorbed \ce{O2} molecules, Bader charge of the O atoms, and net magnetization of the Ni atoms in the phases that appear in the phase diagrams at temperatures $T=300\pm{}100$~K.  All data based on SCAN+rvv10+U calculations. \vspace{0.5\baselineskip}}
\begin{tabular}{*6c}
\toprule
\multicolumn{1}{c}{Facet} & \multicolumn{1}{c}{Phase} & Bond length, \AA{} & \multicolumn{2}{c}{Bader charge, $e^{-}$} & Net magnetization, $\frac{1}{2} \mu_B$\\
\cmidrule(lr){3-3}\cmidrule(lr){4-5}\cmidrule(lr){6-6}
  &   & Surface O-O bonds & Surface O-O bonds & Rest/All O atoms & Ni atoms\\
\midrule
                (001)  & 6         &  1.34 &  -0.45  & -1.17  & 1.02\\
                       & 7         &  1.30 &  -0.32  & -1.13  & 0.95\\
                       & 3         &  -    &  -      & -1.05  & 0.73\\
                       & 8         &  -    &  -      & -0.99  & 0.61\\
                       & 5         &  -    &  -      & -0.92  & 0.49\\
                       & 13        &  -    &  -      & -1.04  & 1.10\\
                       & $13^{*}$  &  -    &  -      & -1.18  & 1.24\\
\midrule
             (104)     & 14        &  -    &  -      & -1.12  & 0.88\\
                       & 15        &  -    &  -      & -1.04  & 0.68\\
                       & 17        &  -    &  -      & -0.98  & 0.84\\
                       & 21        &  -    &  -      & -1.03  & 0.89\\
                       & 22        &  -    &  -      & -1.07  & 0.95\\
                       & 23        &  -    &  -      & -1.15  & 1.08\\
\bottomrule
\label{stbl:bader-MAG-bond-length-info}
\end{tabular}
\end{table}


\begin{table}[H]
	\centering
	\caption{Percentage of lithium and oxygen in the active regions of the surface slab models for the additional phases occurring in the \ce{LiNiO2(001)} surface phase diagrams constructed from PBE+U energies (\textbf{Figure~\ref{fig:reversible-001-pd}}). \vspace{0.5\baselineskip}}
    \small
	\begin{tabular}{lccc}
      \toprule
      & \multicolumn{3}{c}{\textbf{Phase}} \\
      \cmidrule{2-4}
      & \textbf{24} & \textbf{25} & \textbf{26}  \\
      \midrule
      Surface Li (\%) & 100 & 87.5 & 62.5 \\
      Surface O (\%)  & 100 & 100  & 50.0 \\
      \bottomrule
	\end{tabular}
	\label{stbl:addtitional-coverage-percent}
\end{table}

\subsection{Supplementary figures}

\begin{figure}[H]
	\centering
	\includegraphics[width=0.5\textwidth]{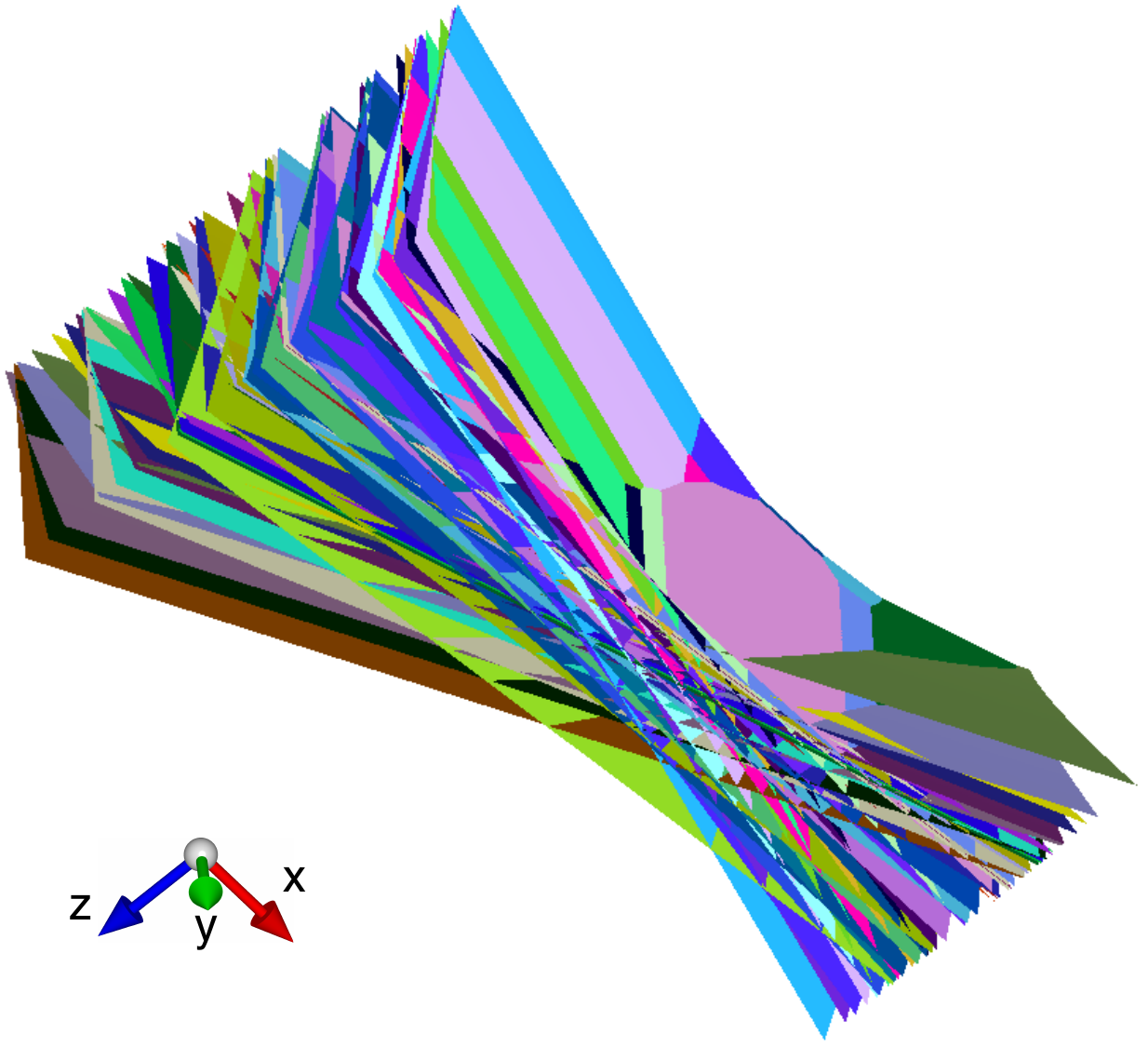}
	\caption{Visualization of the free-energy planes of the \ce{LiNiO2(001)} surface in three dimensions ($x$, $y$, and $z$ coordinates correspond to the potential in V, temperature in K, and energy in eV, respectively).  The surface phase diagram is obtained via projection into a plane. }
	\label{fig:3D-surface-pd-demo}
\end{figure}


\begin{figure}[H]
	\centering
	\includegraphics[width=0.8\textwidth]{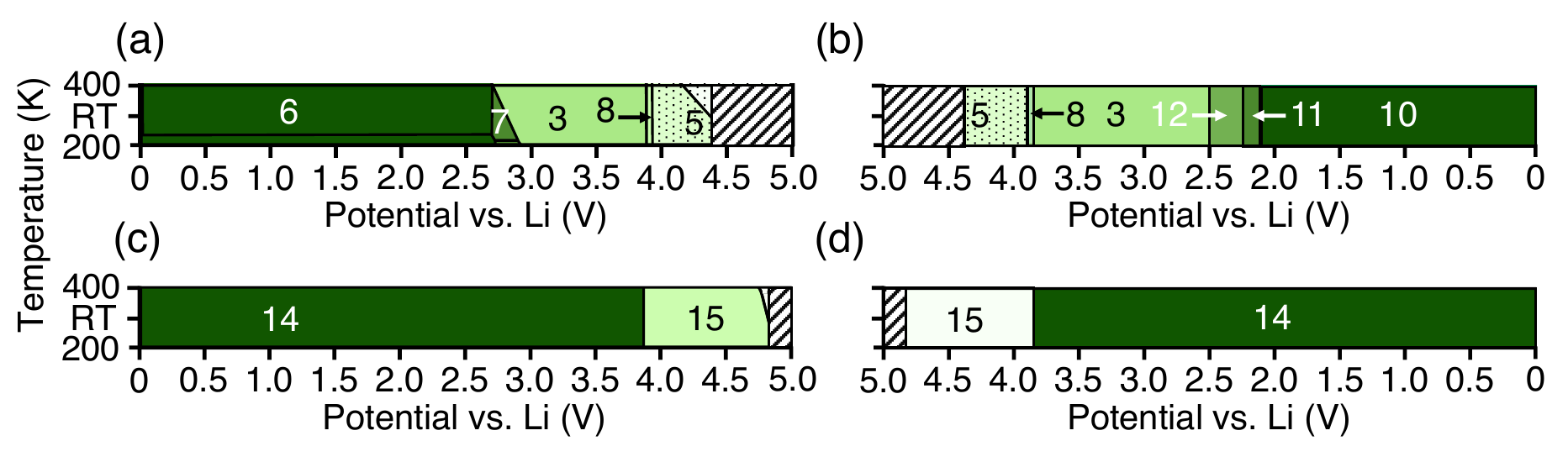}
	\caption{Predicted surface phase diagrams of the \textbf{(a,b)}~\ce{LiNiO2(001)} and \textbf{(c,d)}~\ce{LiNiO2(104)} surfaces during the first charge-discharge cycle near room temperature with lower voltage cutoffs 4.35~V and 4.8~V, respectively.}
	\label{fig:cycling-PDs-diff-cutoff}
	\vspace{-\baselineskip}
\end{figure}

\vspace*{\fill}
\newpage


\begin{figure}[]
	\centering
	\includegraphics[width=0.8\textwidth]{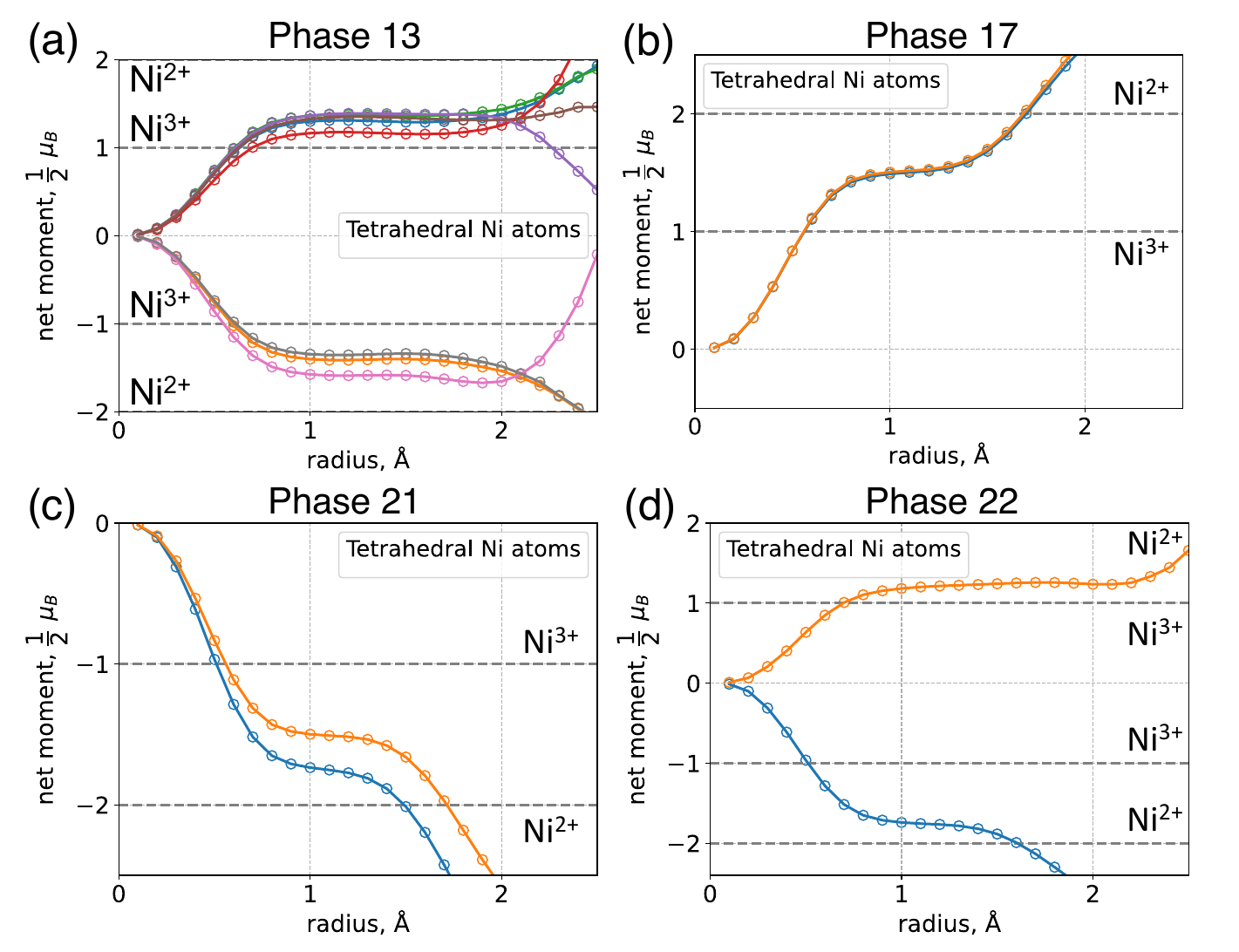}
	\caption{Integrated charge-density differences as a function of integration radius (\AA{}) around Ni atoms in the tetrahedral sites of \textbf{(a)}~phase~13, \textbf{(b)}~phase~17, \textbf{(c)}~phase~21, and \textbf{(d)}~phase~22 as predicted by  \textbf{SCAN+rVV10+U} calculations.}
	\label{fig:spin-integration}
\end{figure}

\vspace{-\baselineskip}

\enlargethispage{\baselineskip}
\begin{figure}[H]
	\centering
	\includegraphics[width=0.6\textwidth]{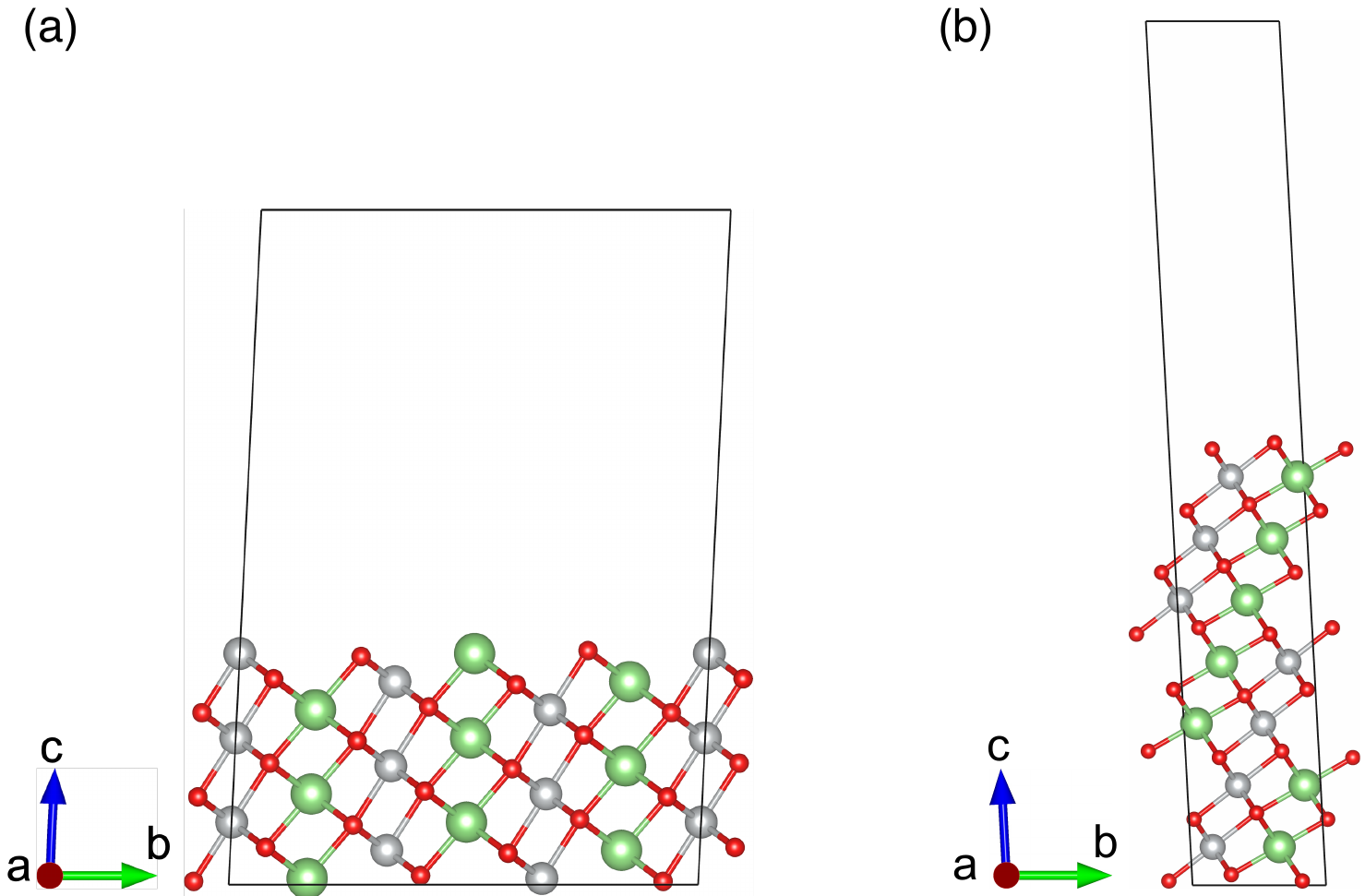}
	\caption{Schematics of the fully lithiated \textbf{(a)} (010) and \textbf{(b)} (012) surface facets.}
	\label{fig:010-012-surface-facets}
\end{figure}

\end{document}